\documentclass[11pt,eqsecnum]{article}

\usepackage{amsmath, amssymb}
 \usepackage{epsfig}
\usepackage{mathrsfs}
\usepackage{psfrag}
\usepackage{graphicx}
\usepackage{epstopdf}

\usepackage{color}

\newcommand \be{\begin{eqnarray}}
\newcommand \ee{\end{eqnarray}}

\newcommand{\ba}{\begin{eqnarray}}
\newcommand{\ea}{\end{eqnarray}}

\newcommand{\refeq}[1]{Eq.~(\ref{#1})}

\def\tdo  {\Omega}
\def\l {\lambda}
\def\sl{\sqrt{\l}}
\def \td {\tilde}
\def \td {\tilde}
\def\sn{{\rm sn}}
\def\cn{{\rm cn}}
\def\dn{{\rm dn}}

\def \EE {{\mathbb{E}}}
\def \KK {{\mathbb{K}}}

\def\no{\nonumber}
\def\d{\partial}
\def\D{{\cal D}}

\def\p{\phi}
\def\b{\beta}
\def \g{\gamma}
\def\o{\omega}
\def\S{{\cal S}}
\def\r{\rho}
\def\a{\alpha}
\def\k{\kappa}
\def\E{{\cal E}}

\def\s{\sigma}

  \def\L{{\cal L}}

 \def\tg{\tilde{\g}}

\def \adss  {$AdS_5 \times S^5$}
\def \sql {\sqrt{\lambda}}
\def\O{{\cal O}}
\def\Om{\Omega}
\def\La{\Lambda}

\begin{document}
\renewcommand{\thefootnote}{\arabic{footnote}}
 
\def \foot {\footnote}
\def \bi{\bibitem}

\def \tr {{\rm tr}}
\def \ha {{1 \over 2}}

\def \ci{\cite}
\def \N {{\mathcal N}}
\def \const {{\rm const}}
\def \t {\tau}
\def\S{{\mathcal S} }
\def \nn {\nu}
\def \XX {{\rm X}}

 \def \vp {\varphi} \def \bs {\bar \s }
\def \k {\kappa}
\def\foot{\footnote}
\def \four{{\textstyle {1\ov 4}}}
 \def \third { \textstyle {1\ov 3
}}
\def\det{\hbox{det}}
\def \ci {\cite}
\def \ov {\over}

\def \bp {\begin{pmatrix}}  \def \epm {\end{pmatrix}}
\def \ha {{\textstyle{1 \ov 2}}}

\def \bi {\bibitem}
\def \la {\label}
\def \Tr  {{\rm Tr}}

\def \T {{\cal T}}
\def \l {\lambda}
\def\foot{\footnote}
\def \tl  {{\tilde \l}}
\def \sql {{\sqrt \l}}
\def \adss {$AdS_5 \times S^5$\ }
\newcommand{\rf}[1]{(\ref{#1})}

\def \bp {\begin{pmatrix}} 
 \def \emp {\end{pmatrix}}
 
 \def \dett  {{\det}}
 
\def \qr {{\hat \rho}}
\def \const {{\rm const}}
\def \bea{\begin{eqnarray}}
\def \eea{\end{eqnarray}}
\def \no {\nonumber}
\def \ov {\over}
\def \tr {{\rm Tr}}
\def \g {\gamma}
\def \tm {\mbb{T}}
\def \ha {\fr{ 1}{ 2}}
\def \half {\fr{ \trm{1}}{\trm{2}}}
\def \s {\sigma}
\def \vp {\varphi}
\def \td {\tilde}
\def \z {\zeta}
\def \H {{\rm H}} 
\def \Tr {{\rm Tr}}
\def \ep {\epsilon}
\def \bp {\begin{pmatrix}} 
 \def \emp {\end{pmatrix}}
\def \ef {\end{document}}
\def \del {\partial}
\def \G {\Gamma} \def \ha { { 1 \ov 2}}  \def \tg  {\td \Gamma} \def \m {\mu}
 \def \tdb {\bar } 
\def \lm {Lam\'e\ }
 \def \tdb {\bar } 

\overfullrule=0pt
\parskip=2pt
\parindent=12pt
\headheight=0in \headsep=0in \topmargin=0in \oddsidemargin=0in

\vspace{ -3cm} \thispagestyle{empty} \vspace{-1cm}
\begin{flushright} AEI-2009-127
\end{flushright}
 \vspace{-1cm}
\begin{flushright} Imperial-TP-AT-2010-01
\end{flushright}
\begin{center}
{\Large\bf
Exact computation of  one-loop  correction to  energy\\ of spinning
folded string in $AdS_5\times  S^5$
 }

 \vspace{0.8cm} {
  M.~Beccaria$^{a,}$\footnote{beccaria@le.infn.it}, G.~V.~Dunne$^{b,}$\footnote{dunne@phys.uconn.edu}, V.~Forini$^{c,}$\footnote{forini@aei.mpg.de} ,
 M.~Pawellek$^{d,}$\footnote{Michael.Pawellek@physik.uni-erlangen.de} and A.~A.~Tseytlin$^{e,}$\footnote{Also at
 Lebedev  Institute, Moscow.\ \
  tseytlin@imperial.ac.uk
 }}\\
 \vskip  0.5cm

\small
{\em
$^{a}$
Physics Department, Salento University and INFN, 73100 Lecce, Italy 

 \vskip 0.05cm
$^{b}$ Department of Physics,  University of Connecticut,
Storrs CT 06269-3046, USA

     \vskip 0.05cm
$^{c}$Max-Planck-Institut f\"ur Gravitationsphysik,  Albert-Einstein-Institut \\
Am M\"uhlenberg 1, D-14476 
Potsdam, Germany

\vskip 0.05cm
$^{d}$  Institut f\"ur Theoretische Physik III, Universit\"at Erlangen-N\"urnberg,\\ 
Staudtstr.7, D-91058 
Erlangen, Germany 

\vskip 0.05cm
$^{e}$  The Blackett Laboratory, Imperial College,
London SW7 2AZ, U.K. }

\normalsize
\end{center}

 \vskip 0.8cm

 \begin{abstract}
We consider the 1-loop correction to the energy of 
 folded spinning string solution in the $AdS_3$  part of $AdS_5 \times S^5$.
The classical string solution is expressed in terms  of elliptic  functions so an  explicit computation  of the corresponding fluctuation  determinants 
for generic values of the spin appears to be   a non-trivial problem. 
We show how it can be solved exactly  by using the static gauge 
expression for the string partition function (which we demonstrate to be equivalent to the conformal gauge one) and  observing that  all the corresponding 
second order fluctuation operators can be put into the standard (single-gap) Lam\'e  form. 
We systematically derive the small spin and large spin expansions
of the resulting expression for the string energy  and comment on some of their
applications. 
\end{abstract}

\newpage

\section{Introduction}

Classical  
string solutions in $AdS_5 \times S^5$, or non-topological
solitons of the string sigma model,
play an important guiding role  in the study of gauge-string duality.
One of the  basic examples is the folded  spinning  string  in AdS 
\ci{gkp,dev}. The classical string energy is a non-trivial  function 
of the spin, interpolating between the 
flat-space regime, $ E \sim \sqrt{ S}+...$,
for small spin  and  the scaling AdS regime,   $E=S + a  \ln S + ...$,
for large spin. The    dual gauge-theory interpretation  
of the latter was suggested in \ci{gkp} and since then 
was used and verified in many papers.

The form of this  spinning string 
solution is determined by an elliptic sn function
(a solution of the sinh-Gordon equation).  Computing the 
quantum correction to its
energy in \adss  string theory is thus a non-trivial problem,  first  addressed in \ci{ft1}.
In \ci{ft1}  the 1-loop correction to the energy  $E_1(S)$   was expressed in terms of 
determinants of  the bosonic and fermionic  fluctuation  operators  with elliptic-function 
potentials. It was explicitly computed only in the  large-spin limit when  the solution
 simplifies  drastically (elliptic function potentials become constant). 
Recently,   attempts were made  to compute the first few leading terms in  $E_1(S)$  in the small $S$ \ci{tt} and the large $S$ \ci{bftt}  expansions.

The aim of the present paper is to solve the problem addressed in \ci{ft1}, i.e. to present   the  general  analytic expression 
for the  1-loop correction $E_1(S)$ for an arbitrary value  of the  spin. 
This would enable us  to  systematically expand $E_1$ in the small $S$ or  large $S$ limits.  

The study of the large $S$   expansion of string energy 
is important for several reasons, e.g.,  
(i) for comparison  with  the Bethe ansatz predictions (see, e.g., 
\ci{bes,bkk,ck,fz}); (ii) for further verification  of the reciprocity property 
at strong coupling \ci{bk,bftt}; (iii) for 
understanding the on-set  of  finite size (exponential or ``wrapping'') corrections in the
  anomalous dimension of the corresponding 
  twist 2  gauge theory operator 
(cf. \ci{ambj,saku,janik})  and the problem of orders of large-spin/large-coupling limits. 
The study of the small $S$ expansion may shed light on  quantum corrections to 
 quantum string states or ``short'' operators  \ci{tt,rt}. 
 
 It is also  interesting to compare the explicit form of the 1-loop 
 string correction derived directly from the string  theory action  
 with the expression coming out of the  approach based on classical  integrability 
 of the string sigma model  \ci{kaz,gr1}. The two should match in general  
 (see \ci{gr2} and refs. there) 
   but  detailed comparison may teach us  important lessons  about  the workings 
 of   the  integrability in the case of cylindrical world-sheet topology. 
 

\

 We shall start in section 2  with a summary  of the basic relations for the 
 classical spinning 
 string solution of \ci{gkp,dev}. Then  in section 3 we shall 
 review the approach of \ci{ft1} to the computation of the 1-loop correction $E_1$ to 
 the string energy. In addition to having  elliptic function potentials
 in the quadratic fluctuation operators, 
 a complication of the conformal gauge expression for the 1-loop partition 
 function of the string sigma model  expanded near this   solution 
 is a mixing of the three $AdS_3$ modes. This mixing is absent 
 in the static gauge  \ci{ft1}, and we go beyond the discussion 
 in \ci{ft1}  by arguing that the  conformal-gauge and the static-gauge 
 expressions are indeed  equivalent (in particular, the string correction in the static gauge is also UV finite). 
 This allows us to use  the static gauge  expression for $E_1$ 
 in which  all 8+8 
 bosonic and fermionic fluctuation modes are decoupled
 as a starting point of our investigation. 
 
 A further crucial observation made  in section 3.3
 is that all second-order fluctuation operators in this stationary 
soliton problem can be put into the standard single-gap  Lam\'e
ordinary differential operator  form on a circle. As discussed in section 4, this  allows  us to compute their 
determinants in an explicit way. In section 4.1 we review  several equivalent forms 
of the 
general expression for the determinant of the second-order ordinary 
differential operator $\O=- \del_x^2 + V(x)$ on a circle:
in terms of the discriminant, in terms of the quasi-momentum,  in terms of the $\zeta$-function or resolvent. 
In section 4.2 we specify  these relations to the case of the 
\lm potential $V= 2 k^2 {\rm sn}^2(x | k^2)$. 
Then in section 4.3  we apply this formalism to the case of the 
fluctuation operators  whose determinants 
appear in the string 1-loop correction $E_1$. 

In section 5 we first demonstrate that the resulting 
expression for $E_1$ is  UV finite  as expected.
 We then check  the equivalence between the conformal gauge and the  static gauge
 expressions for $E_1$  by numerically evaluating the ``mixed'' conformal gauge  determinant
 and comparing it  with its   static gauge counterpart that we found analytically. 
 We also plot $E_1$   and compare it with  its large spin and small spin asymptotics
 derived analytically   in sections 6 and 7. 
In section 6 we  also  check  the reciprocity  constraints 
on the few leading terms in the large spin expansion
of the energy.

Some concluding remarks are made in section 8. 
One natural extension  that we plan 
to address in the future \ci{bdfgpt} is   to repeat the analysis 
of the present paper in the case of the $(S,J)$ 
folded string  solution  with an extra  orbital momentum in $S^5$ \ci{ft1} . 
This problem is more complicated in that even the  static gauge expression for the fluctuation Lagrangian  has  now  two mixed fluctuations
and thus the standard expressions  for the \lm operator determinant cannot be directly applied.


There are several Appendices containing notation and technical details.
In Appendix A we  summarise the   basic definitions for the elliptic functions, and describe the 
Landen transformation used to convert certain fluctuation operators to the Lam\'e form. Appendix 
B describes the Gel'fand-Yaglom numerical method for computing determinants of second-order 
differential operators, including the case of coupled operators. Appendix C contains the details
 of the relevant elliptic function expansions needed for studying the small spin  and the large 
 spin  limits, while in Appendix D we evaluate the leading correction due to the (exponentially 
 suppressed) contributions that we neglect in the main calculation. In Appendix E we consider an
  alternative approach to the expansion of the one-loop energy in the large spin limit.
Appendix F relates our exact results to the perturbative expansion of the associated determinants.


\renewcommand{\theequation}{2.\arabic{equation}}
 \setcounter{equation}{0}
  
\section{Review of  folded spinning string solution in $AdS_3$}


The folded  spinning  string  in $AdS_3$ space  
\be\label{ads3}
 ds^2= -\cosh^2 \rho\ dt^2 + d \rho^2 + \sinh^2 \rho\ d \phi^2
\ee
is a classical closed string solution given by \ci{gkp} 
\be\label{sol}
t= \kappa \,\tau,~~ \quad \phi= \o \,\tau, ~~\quad \rho=\rho(\sigma) =\r(\s+2\pi), 
\ee
where  $\k,\o$ are  constant parameters.
The equation of motion in conformal gauge\footnote{We use Minkowski signature in both target space and
world sheet, so that in conformal gauge $\sqrt{-g}\,g^{ab}=\eta^{ab}={\rm diag}(-1,1)$.}
and its solution  with initial condition $\r(0)=0$ are~\footnote{To construct the full ($2 \pi$ periodic)
  folded closed string solution
  one should  glue together four such functions $\rho(\sigma)$ 
   on $\pi \ov 2$ intervals
  and cover the full $0\leq \sigma \leq 2 \pi$ interval. 
  }
\ba\label{eom}
\rho'^2 &=& \kappa^2\cosh^2\rho-w^2\sinh^2\rho, \\\label{rhop}
\sinh\rho  (\sigma) &=& \frac{k}{\sqrt{1-k^2}}\,\cn(\omega\,\s+\KK \,\,| k^2) \ , \ \ \ \ \ \ \ \ 
\rho' (\sigma) =  \kappa\,{\rm sn}(\o\,\sigma+\KK\,|\,k^2)\ ,
\ea
where $\KK\equiv \KK (k^2)$ is the {\it complete elliptic integral of the first kind} \cite{WW}, with elliptic modulus given by  
 $k\equiv {\kappa\ov \omega}$.\foot{See Appendix A for notation. We adopt 
 here the Abramowitz-Mathematica notation for the modulus of the elliptic functions.}
Here $\rho$ varies from $0$ to its maximal  value $\rho_0$, which  is related to
  the useful parameter $\eta$  or   $k$   by 
\be \label{rho0}
\coth^2 \rho_0 = \frac{\o^2}{\kappa^2}\equiv 1+ \eta\equiv \frac{1}{k^2} \ .
\ee
The periodicity implies an extra condition for the parameters
 \be\label{periodic}
 2\pi=\int_0^{2\pi}d\s=4\int_0^{\rho_0}\frac{d\r}{\sqrt{\k^2\,\cosh^2\r-\o^2\,\sinh^2\r}}
 \ee
 integrating which one finds (see \ref{rho0}) 
 \be\label{kappaomega}
 \k=\frac{2\,k}{\pi}\,\KK  ,~~~~~~~\o=\frac{2}{\pi}\,\KK .
 \ee
The corresponding induced 2-d metric on the $(\t,\s)$ cylinder  and its curvature are 
 \be \la{curv}
 g_{ab}=\r'^2(\s)\,\eta_{ab} \ ,\ \ \ \ \ \ \ \ 
 R^{(2)}=-\frac{\d^2_\s\ln\r'^2}{\r'^2}=-2+\frac{2\,\k^2\,\o^2}{\r'^4}
 \ee
 The two conserved momenta conjugate to $t$ and $\phi$ are 
 the classical energy and the spin 
 \be
 E_0=\sl\,\k\int_0^{2\pi}\frac{d\s}{2\pi}\cosh^2\r\equiv \sl\,\E,~~~~~~~ S=\sl\,\o\int_0^{2\pi}\frac{d\s}{2\pi}\sinh^2\r\equiv \sl\,\S
 \ee
 Using (\ref{eom}) and  (\ref{periodic})  we get 
  the following explicit expressions in terms of the complete elliptic integrals  $\KK=\KK(k^2)$ and $\EE=\EE(k^2)$ (see Appendix A)
 \ba\label{Eclassic}
{\cal E}_0& =& \frac{2}{\pi}\,\frac{k}{1-k^2}\,\EE \ ,   \\\label{Scl}
{\cal S}& =& \frac{2}{\pi}\,\Big(\frac{1}{1-k^2}\,\EE-\KK \Big) \ .
 \ea
To find the energy in terms of the spin one is to solve for $k$ (or $\eta$) in terms of $\S$ 
and then substitute it into the expression for  the energy $\E$. This can be
easily  done in the two limiting cases: 

\noindent 
(i)  large spin or long string  limit: \ \ $\r_0\to\infty$, i.e. $\eta\to0$ or $k\to 1$

\noindent
(ii) small spin or short string limit: \ \   $\r_0\to0$, i.e. $\eta\to\infty$ or  $k\to0$ 

In the   ``long string'' limit  when  the string's  ends are 
close to the boundary of $AdS_5$, the spin
 is automatically large and  the parameter $\eta$ is expanded around zero as 
\ba\label{etasmall}
\eta= \frac{2}{ {\cal S}}-  \frac{  \ln( 8 \pi {\cal S})-3}{ \pi^2 {\cal S}^2}
+...~, ~~~~~~
~~~~~~\eta\ll1
\ea
Substituting this in (\ref{Eclassic}) one obtains for the energy the well known 
logarithmic behavior~\cite{gkp,ft1,bftt}
\ba
{\cal E}_0= \S+\frac{\ln (8\pi\S) -1}{\pi} + \frac{\ln (8\pi\S) -1}{2 \pi^2 \S}+...  \ , \ \ \ \ \ \ \ \ 
\S\gg1 \ ,  \label{Ecl} 
\ea
where the leading $\ln \S$ 
 term is governed by the so-called ``scaling function" (cusp anomaly) and
 the subleading ones can be shown to obey non-trivial reciprocity relations~\cite{bk, bftt}.

In the  ``short string''  limit, when the string is rotating  
 in the small  central ($\rho=0$) region of $AdS_3$, 
 the spin is small and the parameter $\eta$ is large 
\be\label{etalarge}
\frac{1}{\eta}=2\,\S-\frac{1}{2}\,\S^2+\frac{7}{8}\,\S^3-\frac{117}{64}\,\S^4+...~~, ~~~~~~~~~~~~~~\eta\gg1
\ee
This results in the usual flat-space Regge relation~\cite{gkp,ft1,tt}
\be
 \E_0=\sqrt{2\,\S}\Big(1 +\frac{3}{8}\,\S +... \Big) \ , \ \ \ \ \ \ \ \ \ 
\S\ll1 . 
\label{Ec}
\ee
These small and large spin expansions of the classical energy ${\mathcal E}_0$ are shown 
in Figure 1, compared to the exact relation. A similar plot for the one-loop correction
 is provided below, see Figure 8.
\begin{figure}[htb]
\includegraphics[scale=0.75]{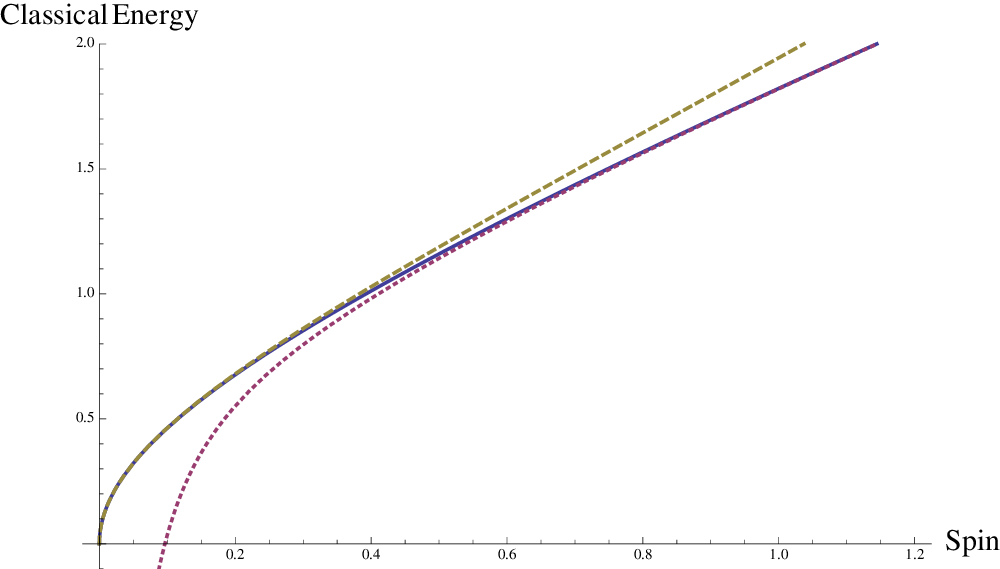}
\caption{Plot (blue, solid curve) of the classical energy ${\mathcal E}_0$ as a function of the spin ${\mathcal S}$, compared with the large spin expansion (red, dotted curve) in (\ref{Ecl}), and the small spin expansion (gold, dashed curve) in (\ref{Ec}). }
\end{figure}

\def \s {\sigma}

\renewcommand{\theequation}{3.\arabic{equation}}
 \setcounter{equation}{0}

\section{One-loop correction to the  spinning string  energy }


As discussed in \ci{ft1}, one can compute the
 leading quantum correction to the energy of this solution 
  by expanding the action to quadratic order in fluctuations 
  near the classical solution
\be\label{genfluct}
\widetilde{I} = -\frac{\sqrt\lambda}{4\,\pi}\int d\tau\,\int_0^{2\,\pi} 
d\sigma\,(\widetilde{\cal{L}}_B + \widetilde{\cal{L}}_F)
\ee
and computing the corresponding partition function expressed in terms of determinants 
of the quadratic fluctuation operators.  
Then (switching  to the euclidean time  $\t\to i\,\t$)
 the 1-loop correction to the energy can 
 be  found from the  2d  effective action $\Gamma$  by dividing over the time interval 
 ($t=\k\,\t$)
\be\label{E1def}
E_1=\frac{\Gamma }{\k\,{\cal T}},~~~~~~~~~~~~{\cal T}\equiv\int d\tau\to\infty,~~~~~~~~~~~\Gamma
=-\ln Z
\ee
where $Z$ is given by the ratio  of the fermionic and bosonic determinants. 

Since the above rigid spinning string solution is stationary,
the coefficients in the fluctuation Lagrangian  do not depend on $\tau$. Then  
the relevant 2-d functional 
determinants  may be reduced  to 1-d  determinants  as in 
\begin{equation}
\ln\det[-\d^2_\s-\d^2_\t+ M^2(\s)]= \T
\int_{-\infty}^{+\infty} {d{\Om}\ov 2\pi} \ \ln \det [-\partial_\s^2+\Om^2+M^2(\s)]
 \    \la{kou}
\end{equation}
i.e. we may introduce $\td \Gamma $ defined by 
\be 
\Gamma= \T \int_{-\infty}^{+\infty}  {d{\Om}\ov 2\pi}\  \td \Gamma   \ . \la{gag} 
\ee

 
\subsection{Conformal gauge}
 
 Following \ci{ft1} one may use either the conformal gauge or the static   gauge to compute the fluctuation 
 Lagrangian  and thus the corresponding 1-loop   partition function. 
 The bosonic fluctuation Lagrangian reads
\begin{eqnarray}\no
&&\td {\cal L}_B^{(\rm conf)}=- \partial_a \td {t} \partial^a \td {t}- \mu_t^2 \td {t}^2 +   \partial_a \td {\phi}
 \partial^a \td {\phi}+ \mu_{\phi}^2 \td {\phi}^2 +\partial_a \td {\rho} \partial^a \td {\rho}+\mu_{\rho}^2 \td {\rho}^2+
  \\
&&~~~~~~~+ 4\, \td {\rho} (\kappa \sinh \rho\ \partial_0 \td {t} - \o\,\cosh \rho\ \partial_0 \td {\phi})+\  \partial_a {\beta}_u \partial^a {\beta}_u +\mu_{\beta}^2 {\beta}_u^2  +
\partial_a {\zeta}_s \partial^a {\zeta}_s \ ,   
 \label{lagconf}\\
&&\mu_t^2= 2 \rho'^2 -\kappa^2, \ \ \quad \mu^2_{\phi}=2 \rho'^2 -\o^2, \ \
\quad \mu^2_{\rho}=2 \rho'^2 -\o^2-\kappa^2,\ \
\quad \mu_{\beta}^2=2 \rho'^2  .    
\label{massesconf}
\eea
Here $\beta_u$ ($u=1,2$) are the two $AdS_5$ fluctuations
transverse to the $AdS_3$ subspace in which the string is moving, while
 $\zeta_s$ ($s=1,...,5$) are fluctuations in $S^5$. 
The three $AdS_3$ fields $(\td t,~\td\r,~\td\p)$ are coupled
so that the corresponding 1-d determinant   in  \rf{kou}  will involve the following $3\times 3$ matrix 
differential operator acting on the  3 fields $X=(\td t,~\td\r,~\td\p)$ \   
(after $\tau \to i \tau, \ \del_\tau  \to  i \Omega $) 
\begin{equation}\label{Q1}
\O_{t\r\phi} =
\bp   \d^2_\s - \Om^2- 2\r'^2+\k^2 &2\Om\,\k\, \sinh\r& 0 \\
 -2\Om\,\k\,\sinh\r &-\d_\s^2+\Om^2+2\r'^2-\o^2& 2\Om\,\o\,\cosh\r\\
  0 & -2\Om\,\o\,\cosh\r & -\d_\s^2+\Om^2+2\r'^2-\o^2-\k^2 \emp
\end{equation}
In addition to the   coefficients being  dependent  on $\s$ according to \rf{rhop},
this mixing   makes   finding  the  determinant of  this operator 
  a   non-trivial problem. 
Taking into account the contribution of the two massless  conformal gauge ghosts,\foot{Here we are implicitly assuming 
that  fluctuation determinants are defined  with flat  rather than (in general curved, for $\rho'\not=\const$) 
induced 2d metric, see  related discussion in \ci{dgt,ft1}.} 
 the 
 bosonic contribution  to $\td \G= \td \G_B + \td \G_F  $ in \rf{gag}   may be written then as  
\bea \la{beb} 
&&\tg^{(\rm conf)}_B = \ha \Big( \ln \det\O_{t\r\phi}    + 2 \ln \det \O_\b  + 3 \ln \det \O_0 \Big)   \ , \\
&& 
 \   \ \ \ \ \ \ \    \   \O_\b =   -\d_\s^2+\Om^2+2\r'^2    \ , \ \ \ \ \ \ \ 
\O_0 =  -\d_\s^2+\Om^2 \ . \la{bee}
\eea
The fermionic part of the quadratic fluctuation Lagrangian   can be put into the form \ci{ft1}
 \be \label{fermlagr}
 \tilde{\cal L}_{_F}  =  2 i ( \bar  \Psi \gamma^a \partial_a \Psi - i\mu_{_F} \bar  \Psi \gamma_3  \Psi) \ ,
 \ \ \ \  \ \ \ \ \ \
 \mu_{_F}= \rho'  \   , \ee
 where $\g_a$ are  2-d gamma matrices  (times a
  unit $8 \times  8$ matrix) and  $\g_3 = {\rm diag} (I, -I)$.  It may be 
  interpreted as describing a system of 4+4    2-d Majorana  fermions with 
  $\sigma$-dependent masses $\pm  \rho' $. 
  Squaring the corresponding Dirac operator, 
  the fermionic contribution to the 2-d  effective action $\tg$ in \rf{gag} 
  can be  written as (see also \cite{tt})
 \bea \label{fer}
&& \tg_F = -\frac{1}{2} \Big( 4  \ln \det \O_{\psi_+} +  4 \ln \det \O_{\psi_-} \Big) \ , 
\\
&& \O_{\psi_\pm}  \equiv    -\d_\s^2+\Om^2  +\mu^2_{\psi_\pm}~,
 ~~~~~~~~~~\mu^2_{\psi_\pm}= \pm \mu'_{_F}  +\mu^2_{_F}    =   \pm \r'' + \r'^2  \ . 
 \label{fermops}
\eea 

 \subsection{Static gauge }
 
 Another approach considered in \ci{ft1}  was to start with the Nambu action and 
 use the same classical solution but impose the 
 static gauge on quantum fluctuations: $\tilde t=\tilde\r=0$.
 \foot{The  classical solution of section 2 
 is also the solution of the  Nambu action as the induced 
 metric is conformally flat. We may define the static gauge by 
 the condition that  $\tau$ and $\sigma$ are such that 
 $t$ and $\rho$   have their classical  values, i.e. do not 
 fluctuate.}
 In this case  the remaining $AdS_3$ mode $\td  \phi$  is decoupled 
 and is described by 
 \bea \la{sta}
&& \td {\cal L}_\phi^{(\rm stat)}=   \partial_a \td {\phi} \partial^a \td {\phi}  + \bar \m^2_\phi  \td {\phi}^2  \ ,
 \\
 && 
  \label{massphi}
\bar \m^2_\p
 = 2 \r'^2  +    \frac{ 2 \k^2 \o^2 }{ \r'^2} \  .    \ee
Then the  static gauge analog of \rf{beb} takes the form 
(the masses of other modes  are the same  as in the conformal gauge 
but there is   no ghost determinant contribution) 
 \bea \la{bes} 
&&\tg^{(\rm stat)}_B = \ha \Big( \ln \det \O_{ \phi}    + 2 \ln \det \O_\b  + 5 \ln \det \O_0 \Big)   \ , \\
 && \la{oph}
 \O_\phi  =   -\d_\s^2+\Om^2+  2 \r'^2  +    \frac{ 2 \k^2 \o^2 }{ \r'^2}     \ , \ee
 while the fermionic   contribution to 1-loop partition function  is the same 
 as in \rf{fermops}. 
 
The advantage of the static  gauge  expression for the  effective   action 
is that here all fluctuation modes are  decoupled  and are described by  elliptic differential 
operators of the same type, $  -\d_\s^2+  V(\s)$. 
On general grounds, one  may expect to find the same  expression for the on-shell 1-loop partition function
 in the two gauges.\foot{For example, in the case of string theory  in    flat 
 target space   
   the  static-gauge Nambu and  conformal-gauge 
    Polyakov  1-loop partition functions     (with nontrivial boundary conditions on a disc)
 are indeed the 
    same \ci{frt}.}
  In this   case one should   get the following relation between the 
  determinants of  the conformal-gauge  operator $ \O_{t\r\phi}$ in \rf{Q1}   and   
  the static-gauge operator  $\O_\phi$   \rf{oph}
 \be \la{eqi}
  \det \O_{t\r\phi} =  \det \O_{\phi}\  ( \det \O_{0})^2  \ , \ee
  where $\O_{0}$ is the massless operator in  \rf{bee}.\foot{Here we assume that 
  one of the massless decoupled  modes   is time-like, like time  mode in $ \O_{t\r\phi}$.}
 A concern  about this equality was raised  in \ci{ft1}  based on the fact that while the conformal gauge 
 1-loop partition function    is UV finite \ci{dgt}, the static  gauge one apparently contains 
   an extra divergent term. Indeed, observing that  
 the  1-loop logarithmic  UV divergence of $\ln Z$  is  given   by  the sum of 
 mass-squared terms  and that  $\bar \m^2_\p$  in \rf{massphi} 
 may be written  in terms of the curvature of the induced metric \rf{curv} as 
 $\bar \m^2_\p =  \sqrt{-g}  (4+  R^{(2)}) $, one finds that  this extra divergence is proportional to 
 $\int d\tau d \sigma  \  \sqrt{-g}  R^{(2)}$. This is proportional  to  the Euler number of the world surface, 
 so one may  suggest \ci{ft1} that it may be cancelled  by the contribution of some  extra  ``topological''  factor
 representing the  ratio of measures in the Polyakov and Nambu  path integrals. 
 
 However, this  extra divergence  actually vanishes in the case of the cylindrical 
 world sheet  appropriate for computing the correction to the energy of a closed string 
 state:  since  $\sqrt{-g}  R^{(2)}$ is a total divergence, 
 as long as the induced metric and thus its curvature are defined to be periodic in $\sigma$, 
 the integral over $\s$ should vanish.\foot{More explicitly,  here  the relevant integral is 
 $ \int^{2\pi}_0  d \s  ( \ln \rho')'' =    [( \ln \rho')']^{2\pi}_0 
 = [{\rho''\ov \rho'}]^{\pi/2}_0 + [ {\rho''\ov \rho'}]^{\pi}_{\pi/2}
+ [{\rho''\ov \rho'}]^{3/2\pi}_{\pi} +  [{\rho''\ov \rho'}]^{2\pi}_{3/2\pi}$,
where $\r''= (\k^2-\omega^2) \sinh \r \cosh \r, \ 
\rho'= \pm \sqrt{ \k^2 \cosh^2  \r - \omega^2 \sinh^2  \r}$. 
While  there is an apparent singularity at the turning points were $\r'=0$, 
 this integral should vanish.   
One may consider using a   suitable regularization of the turning points  to  make this vanishing manifest.}

 As we shall  explicitly show    in Section 5 below, the effective action  in the static gauge 
 given by the sum of   \rf{bes} and \rf{fer} is indeed UV finite. 
Moreover, we shall also  verify the  relation  \rf{eqi}, i.e. demonstrate the equivalence of the conformal gauge and static
 gauge results for the finite  1-loop correction to folded string energy. 
 In \ci{ft1} this equivalence was seen only in the long-string (infinite-spin) limit 
 when the solution   \rf{sol}  approaches  the following  asymptotic solution \ci{ftt} ($\omega \to \k\gg 1 $) 
 \be
 t= \k \tau \ ,\ \ \ \ \  \phi = \k \tau \ , \ \ \ \ \ \ \r= \k \s \ , \ \ \ \ \k=  { 1 \ov \pi} \ln \S  \gg 1   \
 , \la{as} \ee 
 for which  $\r'=\k$=const,  $ R^{(2)}=0$  and the relation \rf{eqi}
 can be easily checked. 
 
 Proving   \rf{eqi} analytically for any $\k$   by direct approach  appears to be non-trivial. 
 One indirect way  to demonstrate \rf{eqi}   is to notice  that since the    corresponding 
 quadratic fluctuation operators   appear  in the  linearized (near folded  string  solution) form of the 
 string equations of motion in the two gauges,   one may be able to relate these operators   by relating the 
  two sets of equations.

 The  conformal-gauge  equations  for small fluctuations  following   from   (\ref{lagconf}) 
 \ba\label{eom1}
&&(\d^2_\t-\d^2_\s)\,\td t+\mu^2_{ t}\,\td t+2\,\k\,\sinh\r\,\d_\t\td \r=0\\\label{eom2}
&&(\d^2_\t-\d^2_\s)\,\td\r+\mu^2_{ \r}\,\td\r+2\,(\k\,\sinh\r\,\d_\t\td  t-\o\,\cosh\r\,\d_\t\td
 \phi)=0\\\label{eom3}
&&(\d^2_\t-\d^2_\s)\,\td \phi+\mu^2_{\phi}\,\td  \phi+2\,\o\,\cosh\r\,\d_\t\td\r=0 \ , 
\ea
  should be supplemented  with  the  conformal gauge conditions (Virasoro constraints) 
\ba\label{v}
&&-\k\,\cosh^2\r\,\d_\t\,{\td  t} +  (\o^2-\k^2)\,\sinh\r\,\cosh\r\,\td\r+\r'\,\d_\s\,\td\r+\o\,\sinh^2\r\,\d_\t\,{\td
 \phi}=0\\\label{vi}
&&-\k\,\cosh^2\r\,\d_\s\,{\td t}+\o\,\sinh^2\r\,\d_\s{\td \phi}+\r'\,\d_\t\,\td\r=0 \ . 
\ea
The latter should  allow one, in principle,   to  eliminate the  two modes (say $\td t$ and $\td \phi$) 
in terms of the third one ($\td \r$), 
getting an effective equation for the latter. 
Since the $\rho$-background does not depend on $\tau$  and since the above  equations 
are linear we may do this elimination at the Fourier mode level, i.e.   replacing 
$\td  t  \to e^{i\,\Om\,\t} \bar t(\s), \  \  \td \phi   \to e^{i\,\Om\,\t} \bar \phi (\s),\ \ 
 \td  \rho    \to e^{i\,\Om\,\t} \bar \rho (\s).$
 Then \rf{v},\rf{vi} imply (changing to euclidean time  notation, i.e.
 $\Omega \to  i \Omega$) 
\be\label{t}
\bar t&=&\frac{\sinh\r}{2 \,\k\,\Om}\ \Big( \del^2_\s   -2\r' \,\d_\s +  \k^2-\o^2 - \Om^2    \Big)\tdb\r\ , \\\label{phi}
\bar \phi&=&-  \frac{\cosh\r}{2\,\o\,\Om}\ \Big(\del_\s^2   +2\r'\,\d_\s  
 - \k^2+  \o^2-  \Om^2  \Big)\tdb\r \ . 
\ea
Substituting this into  the equations of motion (\ref{eom1})-(\ref{eom3})
we find  that one  of them is satisfied automatically while the other two 
become equivalent to  the following fourth-order differential equation for $\td\r$, i.e. 
${\cal O}^{(4)} \tdb \r = 0 $,   where 
\be {\cal O}^{(4)}
\equiv  \del^4_\s +2(\o^2+\k^2- \Om^2-4\r'^2) \ \del^2_\s 
 -   8\r' \r''\  \d_\s +  \k^4+(\Om^2+\o^2)^2 + 2\k^2(\Om^2-\o^2) \ . 
\label{eq4th}\ee
Remarkably, this   operator   can be  factorized as a product of two second-order operators as follows: 
\be
\label{fac}
 {\cal O}^{(4)} =  {\cal O}_1  \cdot  {\cal O}_2  \  ,  \  \ \ \  \ \ \ \ 
 {\cal O}_1
 = (\r')^{-1}\ {\cal O}_\phi\  \r'   \ , \ \ \ \ \ \
  {\cal O}_2  =  {\r'}\ {\cal O}_0 \  {(\r')^{-1}} \ , 
\ee
where $  {\cal O}_\phi $ and ${\cal O}_0$ are the same as the static-gauge operator 
   in \rf{oph} and the massless mode operator in \rf{bee},  respectively. 
The  algebraic ${\r'}$ and $({\r'})^{-1}$ factors  may be  attributed  to a 
change of normalization of the corresponding  fluctuations.\foot{A  similar  discussion could be given at the level of  path integral 
with  the  conformal gauge  condition accounted for  by two delta-functions
(as appropriate if one starts with the Nambu path integral).
The step analogous to \rf{t},\rf{phi}  would then produce an extra $ \det {\cal O}_0$ factor
 as required for balance of degrees of freedom.
 For a discussion of the equivalence 
 of conformal gauge and static gauge partition functions 
 in a simpler case of a homogeneous string solution 
 see also \cite{Frolov:2003tu}.}
 This  way we see how the static gauge operator   ${\cal O}_\phi $
 emerges from the mixed  conformal gauge  fluctuation operator, i.e. provides  support for  
  the relation  \rf{eqi}.

In Section 5 below we shall  verify that the effective action  in the static gauge 
  is indeed UV finite, and we demonstrate 
  the equivalence  \rf{eqi}  between the conformal gauge and the static gauge results for the
   finite  1-loop correction to the folded string energy
  by  computing  the corresponding functional determinants.


\subsection{Lam\'e  form  of the second-order  fluctuation operators   }

To summarize, the  simplest starting point for computing the 
1-loop correction to the folded string energy is thus its representation in terms of 
the 
 1-loop effective action  in the static gauge  given 
by the sum of   \rf{bes} and \rf{fer}, i.e. is expressed in terms of determinants 
 of the following three types of operators defined on 
 periodic functions $ f =(\beta, \phi, \psi_\pm)$ 
 \be \la{ch}
&& \O_f  = -\d^2_\s+ V_f(\s) +\Om^2\ ,~~~~~~~~~~~~~~~~~~~~~~~f (\s)=f (\s+2\pi)
 \ ,\ee
 where (using the form  of the  classical solution  \rf{rhop})
  \bea \label{bebi}
&&   V_\b = 2 \r'^2 \ = 2 \k^2\,\sn^2(\bs\,|\,k^2) \ , \ \ \ \ \ \ \ \ \ \ 
\bs \equiv  \o\,\s+\KK   \ , \\ \la{phb} 
 &&  V_\phi  = 2 \r'^2    +\frac{2\k^2\,\o^2}{\r'^2}\ = 
 2 \k^2\,\sn^2(\bs\,|\,k^2)+2\,\o^2\,{\rm ns}^2(\bs\,|\,k^2) \ , \\  \la{pss} 
  && 
   V_{\psi_\pm} =\r'^2\pm\r''  \ =  \k^2\,\sn^2(\bs\,|\,k^2)\pm\k\,\o\,{\rm cn}(\bs
   \,|\,k^2)\,{\rm dn}(\bs\,|\,k^2)  \ . 
 \eea
   These  potentials are plotted  in Figures  \ref{PotBeta}, \ref{PotPhi}, \ref{PotPsi}
   where we have chosen 
    four particular 
      values of the elliptic  modulus $k = { \kappa \ov \omega}$=0.5, \ 0.9,\ 0.99,\ 0.999.

\begin{figure}[hbtp]
\begin{center}
\includegraphics[scale=1]{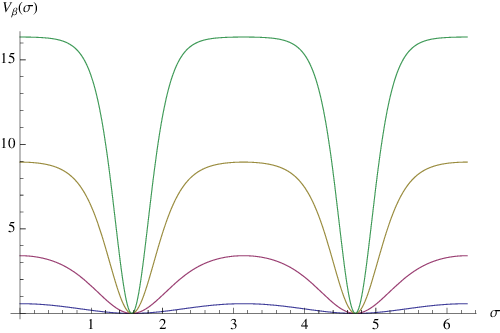}
 \end{center}
 \caption{ Potential $V_\b$ in (\ref{bebi}), 
 for  \(k=0.5, 0.9, 0.99\) and \(0.999\), from bottom to top.}
 \label{PotBeta}
\end{figure}

\begin{figure}[hbtp]
\begin{center}
 \includegraphics[scale=1]{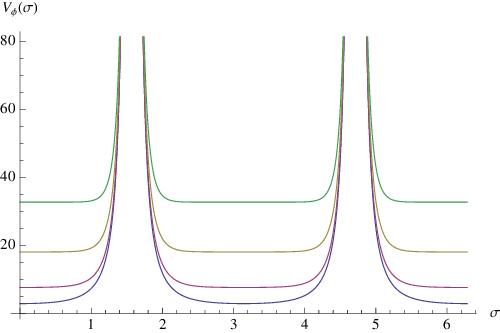}
 \end{center}
 \caption{ Potential $V_\phi$ in (\ref{phb}), for  \(k=0.5, 0.9, 0.99\) and \(0.999\), from bottom to
  top. Note that singularities  appear  at the turning points where 
  \(\sigma=\left(n+\frac{1}{2}\right)\pi\).}
\label{PotPhi}
\end{figure}

\begin{figure}[hbtp]
\begin{center}
\includegraphics[scale=1]{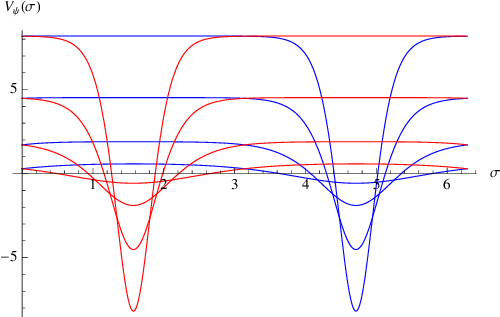}
 \end{center}
 \caption{Potentials $V_{\psi_+}$ (red)  and  $V_{\psi_-}$ (blue) in (\ref{pss}) , for  \(k=0.5, 0.9, 0.99\) and \(0.999\), from bottom to top. Note that $V_\pm(\sigma)$ are identical in from, but are displaced from one another by a half-period $\pi$; they are {\it self-isospectral} \cite{Dunne:1997ia}.}
 \label{PotPsi}
 \end{figure}

It is convenient to introduce  the rescaled spatial variable (cf. \rf{kappaomega}) 
\be 
x \equiv  \o\,\s  = \frac{2\,\KK}{\pi}\,\s   \  , \la{xw} \ee 
and write $\O_{\beta} $ as  (we  ignore a trivial    overall constant factor) 
\ba\label{lamebeta}
\O_{\beta}= 
-\d^2_x+2k^2\,\sn^2(x+\KK\,|\,k^2)+\frac{\pi^2\,{\Om^2}}{4\,\KK^2}  \ ,  
\ea
which is now defined  on the periodic functions $\beta (x) = \beta(x+4\KK)$.
The expression in  (\ref{lamebeta})  is recognized as  being 
a \emph{ Lam\'e} differential operator in the
  \emph{single-gap}  form (which will be  reviewed in the next section). 
  
Remarkably,  all   other fluctuation operators entering the effective action, 
i.e.  $\O_\phi$ and $\O_{\psi_\pm}$, 
 whose structure is apparently much more involved,
   can  also be  cast into  the single-gap Lam\'e form. 
    Their  transformation  has  several steps  involving  rescaling the coordinate
   and   the elliptic modulus 
 a special way, see (\ref{phirelations})-(\ref{phibis}) for  $\O_{\phi}$, and (\ref{landen1})-(\ref{psi-bis})  for  $\O_{\psi_\pm}$.

We can summarize the results as follows: 
each (static gauge) fluctuation  operator  is a  \emph{single-gap Lam\'e operator} 
with  the following  periodic eigenvalue problem  
\be\label{lamegen}
\Big[-\d^2_x+ \ 2\, \bar{k}^2\,\sn^2(x\,|\,\bar{k}^2)+\bar{\Omega}^2\,\Big]\,f_\Lambda(x)=\La\,f_\La(x)\ ,
~~~~~~~~f_\La(x)=f_\La(x+L),
\ee
where $x$ is a  rescaled $\s$ variable with  period $L$ 
and $\bar k$ and $\bar \Omega$ are rescaled modulus and euclidean frequency in \rf{bebi}-\rf{pss}, 
namely,  
\begin{description}
\item{(a)} for the bosonic  operator $\O_\b$:  
\be\label{betarescaled}
x=\frac{2\,\KK}{\pi}\,\s+\KK,~~~~~~\bar k =k\ , 
~~~~~~~~~\bar{\Omega}^2 = \Big(\frac{\pi\,\Om}{2\,\KK} \Big)^2 
,~~~~~~~~~~~~~~~L=4\KK
\ee
\item{(b)} for the bosonic  operator  $\O_\phi$: 
the elliptic modulus is $\td{k}^2=\frac{4k}{(1+k)^2}$ and  
\be\label{phirescaled}
x=\frac{2\,\widetilde{\KK}}{\pi}\,\s+i\,\td \KK'\,,~~~~~~~
\bar k=\td{k} \equiv \frac{2 \sqrt{ k}}{1+k}\ , 
~~~~~~~~\bar{\Omega}^2=
\Big(\frac{\pi\,\Om}{2\,\widetilde{\KK}}\Big)^2+\td{k}^2,~~~~~~L=4\widetilde{\KK}
\ee
\item{(c)} for the fermionic  operators $\O_{\psi_\pm}$:  
\be\label{psirescaled}
x=\begin{cases} 
 \frac{\widetilde{\KK}}{\pi}\,\s+\frac{\widetilde{\KK}}{2},  & \mbox{for}~ \psi_+\\
\frac{\widetilde{\KK}}{\pi}\,\s+\frac{3\,\widetilde{\KK}}{2}, &  \mbox{for}~ \psi_-
\end{cases}
 ~, ~~~~ \bar{k}= \td{k} \equiv\frac{2 \sqrt{ k}}{1+k}\ , 
     ~~~\bar\Omega^2=\Big(\frac{\pi\,\Om}{\widetilde{\KK} }\Big)^2+\td{k}^2,~~~
 ~L=2\widetilde{\KK}
\ee
\end{description}
Here   $\widetilde{\KK}\equiv \KK(\td k^2)$,  and $k' \equiv \sqrt{1 - k^2}$, see Appendix A for notation
and details. 

As we shall discuss  in the next section, the remarkable feature of this
 Lam\'e  spectral problem  (\ref{lamegen}) is that 
 it  can be solved \emph{exactly}, and hence the 
 corresponding determinant  can be  computed \emph{analytically}
(with a result that is  {independent of  constant shifts in the coordinates}).

\renewcommand{\theequation}{4.\arabic{equation}}
 \setcounter{equation}{0}

\section{Determinants of single-gap \lm  operators}

Below we shall   first review  the method  that allows one to 
compute the  determinant of a single-gap \lm  operator without having 
to solve the corresponding spectral
problem explicitly. 
We will  then apply this technique  to the computation of determinants 
of the fluctuation operators discussed in the previous section.

\subsection{Floquet theory of determinants of 2-nd order one-dimensional operators }


Consider the following  eigenvalue problem for an  ordinary differential 
operator, $\mathcal{O}=-\d^2_x+V(x)$, with a  periodic potential
\begin{equation}\label{secondorder}
 \big[-\del_x^2 +V(x)\big]\,f(x) =\Lambda\, f(x) \ , \ \ \ \ \ \ \ 
 V(x+L)=V(x)  \ . 
\end{equation}
For  either periodic  or antiperiodic  boundary conditions on $f(x)$,  we find 
a discrete spectrum of eigenvalues
$\{ \Lambda_n\}$, 
and 
 the associated determinant is then  formally given by 
${\rm Det}{\mathcal O} =\prod_n \Lambda_n$.  

Given  a general potential $V(x)$
 it is of course difficult to find  the eigenvalues, and even given the
  eigenvalues, the infinite product must be regulated. Both difficulties 
  can be overcome in the following way.
Consider  two independent solutions $f_{1,2}(x; \Lambda)$ 
to (\ref{secondorder})   satisfying the conditions 
\begin{eqnarray}
f_1(0; \Lambda)=1 \ , \qquad  \qquad f_1^\prime(0; \Lambda)=0\ , \nonumber\\
f_2(0; \Lambda)=0\ ,  \qquad  \qquad f_2^\prime(0; \Lambda)=1\ , 
\label{norm}
\end{eqnarray}
where $f'= \del_x f$.  
Then the {\it discriminant} $\Delta(\Lambda)$  of the operator $\O$ 
is defined as \cite{magnus}
\begin{eqnarray}
\Delta(\Lambda)=f_1(L; \Lambda)+ f_2^\prime(L; \Lambda)
\label{disc}
\end{eqnarray}
The periodic and the antiperiodic eigenvalues are given by the following 
(in general transcendental) equations:
\begin{eqnarray}\Delta(\Lambda)= \ \begin{cases}+2
 \qquad ({\rm periodic})\\
-2 \qquad ({\rm antiperiodic})
\end{cases}
\label{spectrum}
\end{eqnarray}
Remarkably, the determinant can be computed without knowing these eigenvalues explicitly. 
Indeed, the   Hill determinant, i.e. the ratio of determinants with non-zero $V$ and $V=0$ 
has a simple expression in terms of the discriminant \cite{magnus}:
\begin{eqnarray}
 \frac{\det [-\del_x^2+V(x)-\Lambda]}
{\det[-\del_x^2-\Lambda ]}
&=&\frac{\Delta(\Lambda)-2}{-4\,\sin^2(L\sqrt{\Lambda}/2)} \qquad  ({\rm periodic})  \la{koi} \\
\frac{\det[-\del_x^2+V(x)-\Lambda ]}
{\det [-\del_x^2-\Lambda]}
&=&\frac{\Delta(\Lambda)+2}{4\,\cos^2(L\sqrt{\Lambda}/2)}
\qquad  ({\rm antiperiodic}) \label{hill}
\end{eqnarray}
In what follows we shall  always assume  that determinants 
we consider are normalized  to the  trivial free determinant $\det [-\del_x^2]$, and thus 
omit the resulting $\Lambda-$ and $V$-independent overall constant
(such constants will cancel in the string partition function due to balance of the  degrees of
freedom). 
Then  we may  write the above relations  simply as 
\begin{eqnarray}
\dett_{_{{P, AP}}}[-\del_x^2+V(x)-\Lambda]= 
\begin{cases}
    \Delta(\Lambda) -2                \qquad ({\rm periodic})\\
      \Delta(\Lambda) +2                   \qquad ({\rm antiperiodic})
\end{cases}
\label{hill2}
\end{eqnarray}
It is useful to relate  this representation for the determinant 
to  a familiar physical notion of ``quasi-momentum". By 
 the Floquet/Bloch theory \cite{magnus}, the equation (\ref{secondorder})
  has two independent solutions of the form
$f_\pm (x)=e^{\pm i \, p(\Lambda) \, x}\, \chi_\pm (x)$, 
where $\chi_\pm (x)$ are periodic, so that under translation through one period
 the Bloch solutions $f_\pm(x)$ change by a phase
\begin{eqnarray}
 f_\pm(x+L)=e^{\pm i p(\Lambda)\, L}\ f_\pm(x)
 \label{momentum}
 \end{eqnarray}
 where, by definition, $p(\Lambda)$ is the ``quasi-momentum".
  Then $\Delta(\Lambda)=2\cos (L\, p(\Lambda))$, and 
   we can re-write (\ref{hill2}) in terms of the quasi-momentum
   as follows  \cite{magnus}:
 \begin{eqnarray}
 \dett_{_{P, AP}}[-\del_x^2+V(x)-\Lambda ]=
 \begin{cases}
-4\sin^2\left(\frac{L}{2}\, p(\Lambda)\right)\qquad ({\rm periodic})\\
+4\cos^2\left(\frac{L}{2}\, p(\Lambda)\right)\qquad ({\rm antiperiodic})
\end{cases}
\label{hill3}
\end{eqnarray}
Thus, knowing the quasi-momentum $p(\Lambda)$ amounts to knowing the discriminant and also the determinant.

Another interesting and useful relation is the 
 link between the determinant and the discriminant through 
 the contour integral representation for  the spectral zeta function. 
 For definiteness, let us  consider the case of the 
 periodic boundary conditions. Then  the spectral zeta function is
\begin{equation}\label{ContourZeta}
 \zeta(s)=\frac{1}{2\pi i}\int_{\gamma}\mathrm{d}\Lambda\,\Lambda^{-s}
 \ \frac{\partial}{\partial\Lambda}\ln\left[\Delta(\Lambda)-2\right]=
 \frac{1}{2\pi i}\int_{\gamma}\mathrm{d}\Lambda\,\Lambda^{-s}\ R(\Lambda),
\end{equation}
where the resolvent
\be  \la{reso}
R(\Lambda) = {\Delta'(\Lambda) \ov \Delta (\Lambda)- 2}   \ee
has simple poles exactly at the  values of $\Lambda$ corresponding to the 
points  of the periodic spectrum. 
The contour $\gamma$ in (\ref{ContourZeta}) runs counter-clockwise
above and below the positive real axis enclosing all poles of the resolvent.
Wrapping the contour along the branch cut along the negative real line gives \cite{Kirsten:2004qv}
\begin{equation}
 \zeta(s)=-\frac{\sin(\pi s)}{\pi}\int_0^{\infty}\mathrm{d}\Lambda\,\Lambda^{-s}\ R(-\Lambda).
\end{equation}
According to  the zeta function definition of the functional determinant
\begin{equation}
 \det\left[-\d^2_x+V(x)\right]=e^{-\zeta'(0)}   
\end{equation}
to compute the determinant we need to know 
\begin{equation}
- \zeta'(0)=-\int_0^{\infty}\mathrm{d}\Lambda\frac{\partial}{\partial\Lambda}\ln\left[\Delta(-\Lambda)-2\right]
 =\ln\frac{\left[\Delta(0)-2\right]}{\left[\Delta(-\infty)-2\right]}
\end{equation}
Here we  subtracted the 
 divergent term 
 \(\ln[\Delta(-\infty)-2]\)  by assuming  that we again
 divide by  a ``free''  reference determinant.
 Then  finally \begin{equation}\label{discromega}
 \dett_P\left[-\d^2_x+V(x)\right]=\Delta(0)-2
\end{equation}
Shifting the potential by a constant 
$-\Lambda$, we reproduce  the representation  in (\ref{hill2}).

The important feature of the above expressions is that  the  determinants can be 
calculated in closed form without computing any of the eigenvalues. 
There is yet another way to compute the determinants, known as the
 Gel'fand-Yaglom method~\cite{dunne,kleinert}, which for periodic 
 systems reduces  essentially to a numerical evaluation of the discriminant
   giving the determinants via (\ref{hill2}). This method is described 
  in Appendix B, where we also consider   systems of coupled equations, which will be important for demonstrating explicitly the equivalence of the computation in the conformal and static gauges.



It is useful to illustrate the above general relations on the simple example
of  constant potential 
\be V(x)=m^2, \ \ \ \ \ \ \  x \in (0, L) \ . \ee
Then the two independent solutions in \rf{norm}
are $f_1(x; \Lambda)=\cosh(\sqrt{m^2-\Lambda}\, x)$ and 
$f_2(x; \Lambda)=\sinh(\sqrt{m^2-\Lambda}\, x)/\sqrt{m^2-\Lambda}$. Therefore,
 the discriminant \rf{disc} and the determinants \rf{hill3}
 are 
\begin{eqnarray}
&&\ \ \ \ \   \Delta(\Lambda) =2\cosh(L\, \sqrt{m^2-\Lambda}) \ , \\
&&\dett_{_{P, AP}} (-\del_x^2+m^2-\Lambda)= 
\begin{cases}
4\sinh^2\left(\frac{L}{2}\, \sqrt{m^2-\Lambda} \right)\\
4\cosh^2\left(\frac{L}{2}\, \, \sqrt{m^2-\Lambda} \right)
\end{cases}
\label{constant}
\end{eqnarray}
The quasi-momentum in \rf{momentum}
 here  is $p(\Lambda)=\sqrt{\Lambda-m^2}$, so these 
 relations 
 are consistent with (\ref{hill3}).
  Furthermore, in this case we know  the explicit eigenvalues  ($n\in {\mathbb Z}$):
\begin{eqnarray}
\Lambda_n =
\begin{cases}
m^2+\big(\frac{2n\pi}{L}\big)^2\ \ \ \  \quad \qquad ({\rm periodic})\\
m^2+\big(\frac{(2n+1)\pi}{L}\big)^2 \quad \qquad ({\rm antiperiodic})
\end{cases}
\end{eqnarray}
Then the expressions in (\ref{constant}) also follow 
from the infinite product representations 
 for the sinh and cosh functions, combined with zeta function regularization.

\subsection{Case of single-gap Lam\'e  potential $V(x)=2 k^2\, {\rm sn}^2(x\, |\, k^2)$}

The   important example  which is our main interest here
is provided by the single-gap Lam\'e operator in  \rf{lamegen}, i.e.  
\be\label{lamegen2}
\Big[-\d^2_x+\  2k^2\,\sn^2(x\,|\,k^2)\Big]\,f(x)=\Lambda\,f(x) .
\ee
The two independent Bloch solutions of (\ref{lamegen2}) here are~\cite{WW}
\be\label{solslame}
f_{\pm}(x) = \frac{H(x\pm\alpha)}{\Theta(x)}\,e^{\mp\,x\, Z(\alpha)} \ ,
\ee
where $H, \Theta, Z$ are the Jacobi Eta, Theta and Zeta functions defined in (\ref{jacobidef}), and $ \alpha=\alpha(\Lambda)$ is given implicitly by
\be\label{alphaeq}
{\rm sn}(\alpha\,|\,k^2) = \sqrt{\frac{1+k^2-\Lambda}{k^2}}\ .
\ee
Using the period properties of the Jacobi functions (\ref{periodicityH}) we see that  
\be\label{solperiodic}
f_{\pm}(x+2\KK)=-f_{\pm}(x)\,e^{\mp\,2\,\KK\, Z(\alpha)}\,\equiv\,f_{\pm}(x)\,e^{2i\,\KK\,p(\a)}\ , 
\ee
which defines the quasi-momentum as
\be\label{momentums}
p(\Lambda) = i\, Z(\alpha\,|\,k^2)+\frac{\pi}{2\,\mathbb{K}} \ .
\ee
Therefore, from (\ref{hill3}) we immediately find analytic expressions for the determinants. 
Assuming  the period is  $L=2{\mathbb K}$, we find
\be\label{lamedets}
\det_{_{P, AP}}^{(L=2\KK)}\Big[\!-\d^2_x+2k^2\,\sn^2(x\,|\,k^2)-\Lambda\Big]=
\begin{cases}  
- 4\cosh^2[\KK\,Z(\a | k^2)]    \\ 
 -4\sinh^2[\KK\,Z(\a | k^2)]
\end{cases}
\ee
where the relation between $\Lambda$ and $\alpha$ is given by (\ref{alphaeq}). 
On the other hand, for the 
period $L=4{\mathbb K}$, we find
\be\label{periodicdet}
\det_{P}^{(L=4\KK)}\Big[\!-\d^2_x+2k^2\,\sn^2(x\,|\,k^2)-\Lambda\Big]=
 4\sinh^2[2\,\KK\,Z(\a | k^2)]
\ee
Note that this is the same as   the product 
of the periodic and antiperiodic determinants \rf{lamedets} with  the period $2\KK$, as it should be.

The periodic potential in (\ref{lamegen2}) has the special property that its band spectrum has only a single gap, and is known therefore as a one-gap potential, as illustrated in  Fig. 5. 
The spectrum has  three band edges (which are also the lowest eigenvalues of the periodic 
spectrum of the problem on the interval $4\KK$):
\be
 {\Lambda}_1=k^2\ , \quad  \quad  {\Lambda}_2=1\ , \quad \quad {\Lambda}_3=1+k^2\ .
\label{edges}
\ee
\begin{figure}[hbtp]
\begin{center}
\includegraphics[scale=1.3]{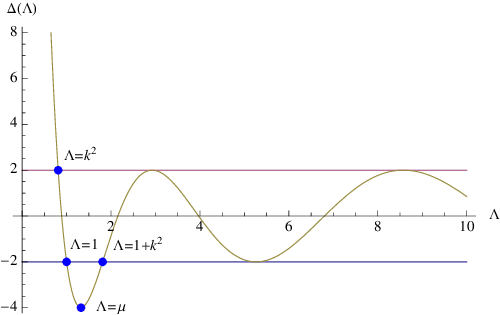}
 \caption{ The discriminant $\Delta(\Lambda)$ for the Lam\'e potential $V(x)=2\,k^2\,\sn^2(x\,|\,k^2)$ with  \(k = 0.9\). The three band edges occur at the points $ {\Lambda}_1, {\Lambda}_2, {\Lambda}_3$ where $\Delta(\Lambda)$ cuts the lines $\pm 2$, while the remainder of the periodic/antiperiodic spectrum consists of points where  $\Delta(\Lambda)$ touches the lines $\pm 2$.}
 \end{center}
 \label{fig:gap}
\end{figure}
One can rewrite the relation between $\Lambda$ and $\alpha$ in (\ref{alphaeq}) 
in terms of the band edges as follows
\be\label{alphaeqbis}
k\, {\rm sn}(\alpha; k^2)=\sqrt{\frac{1}{2}\left({\Lambda}_1+ {\Lambda}_2+ {\Lambda}_3\right) -{\Lambda}}.
\ee
The resolvent is
\begin{eqnarray}
R( {\Lambda})&=&\frac{d}{d{\Lambda}} \,\ln\, [\Delta( {\Lambda})-2]=L\, \frac{d p}{d {\Lambda}}\,
 \cot\big[\frac{L}{2}\, p( {\Lambda})\big] \la{ret} \ . 
\end{eqnarray}
We can also express $dp/d\Lambda$ simply in terms of the band edges:
\be\la{uu}
\frac{dp}{d \Lambda}=i\,\frac{\Lambda-\mu}{2\sqrt{(
 {\Lambda}_1- {\Lambda})( {\Lambda}_2- {\Lambda})( {\Lambda}_3- {\Lambda})}}
\ee
where
\be\la{mam}
\mu=\frac{1}{2} \Big( {\Lambda}_1+ {\Lambda}_2+ {\Lambda}_3 -\langle V\rangle\Big)\ , \qquad
 \langle V\rangle\equiv \frac{1}{L}\,\int_0^{L}V(x)\,dx
\ee
To see this, note that from (\ref{momentum})
\begin{eqnarray}\no
\frac{d p}{d{\Lambda}}
=\frac{d p}{d\alpha}\frac{d \alpha}{d {\Lambda}}=i \,\frac{dZ(\alpha\,|\, k^2)}{d\a}\,\frac{d \alpha}{d {\Lambda}}
=i\,\left(1-k^2 {\rm sn}^2(\alpha\,|\,k^2)-\frac{{\mathbb E(k^2)} }{{\mathbb K(k^2)} }\right)\frac{d \alpha}{d {\Lambda}}  
\end{eqnarray}
where we have used the definition (\ref{zed}) of the Zeta function. Also, from  (\ref{alphaeqbis})  we have
\begin{eqnarray}\no
\frac{d \alpha}{d{\Lambda}}=\frac{1}{2k^2\,{\rm dn}(\alpha\,|\,k^2)\, {\rm cn}(\alpha\,|\, k^2)\, {\rm sn}(\alpha\,|\, k^2)} =\frac{1}{2 \sqrt{( {\Lambda}_1- {\Lambda})( {\Lambda}_2- {\Lambda})( {\Lambda}_3- {\Lambda})}}.
\end{eqnarray}
Finally, for the potential $V(x)=2k^2\,\sn^2(x\,|\,k^2)$  
we find for $L=2\KK$
\be\label{Velliptic}
\langle V\rangle=\frac{1}{L}\int_0^{L}dx\,2k^2\,\sn^2(x\,|\,k^2)=2\Big(1-\frac{\EE(k^2)}{\KK(k^2)}\Big) \ , 
\ee
and the same   for $L=4\KK$. 
Thus, taking into account (\ref{edges}), we get
\be\label{mu}
\mu=k^2+\frac{\EE(k^2)}{\KK(k^2)}  \  .
\ee
Thus we 
 obtain a compact expression for the resolvent of
  the single gap Lam\'e potential with period $L$ 
  in terms of the quasi-momentum $p(\Lambda)$ and the band edges as
\be\label{resolvent}
R(\Lambda)= \frac{L}{2}\,\frac{\Lambda-\mu}{\sqrt{( {\Lambda}_1- {\Lambda})( 
{\Lambda}_2- {\Lambda})( {\Lambda}_3- {\Lambda})}}\,\coth\Big(\frac{L\,p(\Lambda)}{2 i}\Big)  \ . 
\ee

\subsection{Results for determinants of static-gauge  fluctuation operators}

Let us now apply the above results  to the case  of the 
 fluctuation operators defined by  (\ref{lamegen})-(\ref{psirescaled}). 
The results  in Eqs. (\ref{alphaeq}) and  (\ref{lamedets})-(\ref{periodicdet}) are
 actually all that we need  in order
 to write down \emph{exact} analytic expressions for the determinants of these 
   operators. The analytically
  known eigenvalues or band edges can be obtained from (\ref{edges}) with the appropriate shifts 
  ($\La_i\to\La_i-\bar\Omega^2$) and rescalings,
   and an  analogous  procedure applies to the 
  corresponding resolvents in (\ref{resolvent}).

The results  can be summarized as follows. 

\begin{description}
\item (a)
for the \emph{$\beta$ operator}, in view of (\ref{betarescaled}), the determinant reads 
\bea\label{detbeta}
&&\det ~\O_\beta(\Omega)=4\sinh^2\left[2\,  {\mathbb K} \, { Z}(\alpha_\b\,|\, k^2)\right] \ , \\
&&{\rm sn}(\alpha_\b; k^2)=\frac{\sqrt{1+k^2+(\frac{\pi\,\Omega}{2\KK})^2}}{k} \la{sne}\ . 
\eea
The  band edges are obtained from (\ref{edges}) by shifting and  rescaling 
\be
\bar\La_i=\Big(\frac{2\KK}{\pi}\Big)^2(\La_i+\bar{\Omega}^2)\equiv\,\Omega_i^2+\Om^2\ ,~
~~~~~~~~~~~~~\bar{\Omega}^2=\Big(\frac{\pi\,\Om}{2\,\KK}\Big)^2
\ee
where the rescaled $\Lambda_i$ have been defined as ``characteristic frequencies'' $\Omega_i^2$.
\begin{table}\label{tab:spec}
\begin{center}
\begin{tabular}{c||c|c|c}
  & \(\beta\) & \(\phi\) & fermions  \\
 \hline\hline & & & \\ \(\bar\Lambda_1=\Om_1^2+\Om^2\) & \(\kappa^2+\Om^2\) & \(\Om^2\) & \(\Om^2\) \\ & & & \\
\(\bar\Lambda_2=\Om_2^2+\Om^2\) & \(\omega^2+\Om^2\) & \((\omega-\kappa)^2+\Om^2\) & \(
 \frac{1}{4} (\omega-\kappa)^2+\Om^2\) \\ & & & \\
\(\bar\Lambda_3=\Om_3^2+\Om^2\) & \(\omega^2+\kappa^2+\Om^2\)& \((\omega+\kappa)^2+\Om^2\) &
 \(  \frac{1}{4}(\omega+\kappa)^2+\Om^2\) \\ & & & \\
\(\bar{\mu}\) & \(\kappa^2+\omega^2\frac{\EE}{\KK}\) & \(-(\omega^2-\kappa^2)+2\omega^2\frac{\EE}{\KK}\) & 
 \(-\frac{1}{4}(\omega^2-\kappa^2)+\frac{\omega^2}{2}\frac{\EE}{\KK}\) \\
\hline
\end{tabular}
\end{center}
\caption{The lowest (analytically known) eigenvalues of  the fluctuation operators}
\end{table}
One thus gets  the eigenvalues in the first column of Table 1 
that can now be re-expressed in terms of the parameters of the classical solution
\ba \no
\{\bar\La_1,\bar\La_2,\bar\La_3\}&=&\Big(\frac{2\KK}{\pi}\Big)^2\,\Big\{k^2+\bar{\Omega}^2,1+\bar{\Omega}^2,
1+k^2+\bar{\Omega}^2\,\Big\}   \\\label{omegai}
&\equiv&\,\Big\{\k^2+\Om^2,\ \o^2+\Om^2,\ \k^2+\o^2+\Om^2\Big\}.
\ea

\item (b)
For the \emph{$\phi$ operator} in  (\ref{phirescaled}) 
we have $\tilde{k}^2=\frac{4k}{(k+1)^2}$, and  thus 
\bea\label{detphi}
&&\det\O_\phi( \Om)=4\sinh^2\big[2\,  {\widetilde{\mathbb K}} \, { Z}(\alpha_\p\,|\, \td{k}^2)\big]\ , \\ 
&&{\rm sn}(\alpha_\p\,|\, \td{k}^2)=\frac{\sqrt{1 +(\frac{\pi\,\Om}{2\widetilde{\KK}})^2}}{\td{k}} \ . \la{snm}
\ee 
The  band edges are obtained from (\ref{edges}) by shifting and rescaling  
\be
\bar\La_i=\Big(\frac{2\widetilde{\KK}}{\pi}\Big)^2 (\La_i+\bar{\Omega}^2)\equiv\,\Omega_i^2+\Om^2\ ,~~
~~~~~~~~~~~~~~~~\bar\Omega^2=\Big(\frac{\pi\,\Om}{2\,\widetilde{\KK}}\Big)^2+\td{k}^2 \ , 
\ee
getting thus  the eigenvalues in the second column of Table 1. 

\item (c)
For the  \emph{ $\psi_\pm$   operators} in (\ref{psirescaled}) 
  the elliptic parameter is $\tilde{k}^2=\frac{4k}{(k+1)^2}$, and we get 
\bea\label{detpsi}
&&\det\O_\psi(\Om)=-4\cosh^2\big[ \widetilde{{\mathbb K}} \, {Z}(\alpha_\psi\,|\, \td{k}^2)\big] \ , \\
&&{\rm sn}(\alpha_\psi\,|\, \td{k}^2)=\frac{\sqrt{1 +(\frac{\pi\,\Om}{\widetilde{\KK}})^2}}{\td{k}} \ . \la{swe}
\eea 
Since the  determinant is  independent of
 constant shifts of  coordinates like the one in \rf{psirescaled}
 (see, e.g.,   Appendix B)
the  expressions for the determinants of  $\O_{\psi_-}$ and
  $\O_{\psi_+}$ are the same and therefore  we will not distinguish them in what follows. 
The  band edges follow  from (\ref{edges}) 
\be
\bar\La_i=\Big(\frac{\widetilde{\KK}}{\pi}\Big)^2 (\La_i+\bar{\Omega}^2)\equiv\,\Omega_i^2+\Om^2,~~~~
~~~~~~~~~~~~~~~~~~~~\bar\Omega^2=\Big(\frac{\pi\,\Om}{\widetilde{\KK}}\Big)^2+\td{k}^2  
\ee

\end{description}

\renewcommand{\theequation}{5.\arabic{equation}}
 \setcounter{equation}{0}

\section{Exact expression for one-loop  
correction to string  energy }



As follows from the above   discussion (see \rf{E1def},\rf{bes},\rf{fer},\rf{detbeta},\rf{detphi},\rf{detpsi})
 the  1-loop correction to the 
energy of the folded spinning string  may be written as 
\begin{equation}\label{final}
E_1=
-\frac{1}{4\pi \k }\int_{-\infty}^{\infty}\mathrm{d}\Om\ \ln\frac{\det^8
\mathcal{O}_\psi}{ \det\mathcal{O}_{\phi}  \ \det^2\mathcal{O}_{\beta}\,
\,\det^5\O_0},
\end{equation}
where   $\k={2k \ov \pi}\,\KK, \ \ \KK= \KK(k^2)$  (see \rf{kappaomega}) 
and the determinants as functions of $\Omega$    have the following explicit expressions\footnote{The determinant  of the massless   operator $\O_0$ 
 is found by taking  the regularized infinite product of its eigenvalues $\lambda_n=n^2+\Omega^2$. }
\ba\label{detbeta2}
\det\O_\beta &=&~4\sinh^2\left[2\,  {\mathbb K} \, {Z}(\alpha_\b\,|\, k^2)\right]
\qquad {\rm where} \qquad
{\rm sn}(\alpha_\b\,|\,k^2)=\frac{\sqrt{1+k^2+(\frac{\pi\,\Om}{2\KK})^2}}{k}\\\label{detphi2}
\det\O_\phi &=&~4\sinh^2\left[2\,  \widetilde{{\mathbb K}} \, { Z}(\alpha_\p\,|\, \td{k}^2)\right]
\qquad {\rm where} \qquad
{\rm sn}(\alpha_\p\,|\, \td{k}^2)=\frac{\sqrt{1 +(\frac{\pi\,\Om}{2\widetilde{\KK}})^2}}{\td{k}}\\\label{detpsi2}
\det\O_\psi &=&-4\cosh^2\left[\widetilde{ {\mathbb K}} \, {Z}(\alpha_\psi\,|\, \td{k}^2)\right]
\qquad~ {\rm where} \qquad
{\rm sn}(\alpha_\psi\,|\, \td{k}^2)=\frac{\sqrt{1 +(\frac{\pi\,
\Om}{\widetilde{\KK}})^2}}{\td{k}}\\
\label{dettrivial}
\det\O_0 &=&~4\sinh^2\left[\pi\,\Om\right]
\ea
The computation of  $E_1$  is thus reduced to inverting the transcendental equations 
for $\alpha_\b, \alpha_\phi,\alpha_\phi$, finding the corresponding values of $Z$-function \rf{intzeta}.
The integral is then a function of $k={\k \ov \omega}$ and $\Omega$.\foot{Note that in the limit
when $k=0$, i.e. potentials vanish all determinants take the same value 
as $\det\O_0$, i.e. $E_1$ in \rf{final} vanishes.}
Doing the integral over $\Omega$ we then end up with a function of $k$ only  or the 
spin \rf{Scl}.  It is straightforward to evaluate the $\Omega$ integral numerically, as discussed below.

\subsection{UV finiteness}

Let us first check that the resulting expression for $E_1$ is  indeed UV finite, 
i.e. the integral over $\Om$ is convergent at infinity. 
The large $\Omega$ behavior of the determinant factors in (\ref{final}) can most easily be extracted from the general large $\Omega$ behavior of the associated resolvents. Changing variable from $\Lambda$ to $-\Omega^2$, we define
\ba
{\mathcal R}(\Omega)\equiv -2\,\Omega\, R(-\Omega^2)
\label{newR}
\ea
Then we find from (\ref{resolvent}) that
the general structure of the expansion is
\bea \label{Resolventexpand}
 &&\mathcal{R}(\Om)=r_0+\frac{r_1}{\Om^2}+\frac{r_2}{\Om^4}+\mathcal{O}(\Om^{-6})\ ,~~~~~~~~~\Om\to\infty
\\ 
&& r_0 =2\pi\  ,~~~~~~~~~~~~~~~~~
 r_1 = 2\pi\big[\bar\mu-\frac{1}{2}(\bar\La_1+\bar\La_2+\bar\La_3)\big]=2\pi\,\langle V\rangle
 \eea
Therefore, the large $\Omega$ behavior of the log determinant is
\begin{equation}
 \ln\det\mathcal{O}= {r_0\,\Omega}  -  { r_1\ov \Om}+\mathcal{O}(\Om^{-3 }),~~~~~~~~~\tdo\to\infty
\end{equation}
Using the corresponding values of  \(\bar\mu\) and \(\L_i\) of the three non trivial
 fluctuation modes given  in  Table 1, we find [we have also used the elliptic identities (\ref{usefulelliptic1})--(\ref{usefulelliptic2})]:
\begin{eqnarray}
 \ln\det\mathcal{O}_{\beta} &=& 2\pi\,\Om  + 4\,\omega\,(\KK-\EE)\,\Om^{-1}+\mathcal{O}(\Om^{-3}),\la{bej}\\
 \ln\det\mathcal{O}_{\phi} &=& 2\pi\,\Om +  8\,\omega\,(\KK-\EE) \, \Om^{-1}+\mathcal{O}(\Om^{-3}),\la{phj} \\
 \ln\det\mathcal{O}_\psi &=& 2\pi\,\Om  +  2\,\omega\,(\KK-\EE)\, \Om^{-1}+\mathcal{O}(\Om^{-3}) \ , \la{psj}\\ 
  \ln\det\mathcal{O}_{0} &=& 2\pi\,\Om+\mathcal{O}(\Om^{-3})   \ . 
\end{eqnarray}
The leading (quadratically divergent) 
 terms cancel in \rf{final}  due to the balance of world-sheet degrees of freedom in \rf{final}.
  The 
 subleading  (logarithmically divergent) terms 
 also   cancel  in the combination appearing in \rf{final}, 
 $ (  2 \times 4 + 8 -  8 \times 2)  (\KK-\EE)\, \Om^{-1}=0$. 
We thus confirm that the static gauge result for the 1-loop energy
 is indeed UV finite, as was argued in section  3.2.



\subsection{Equivalence  between  the static gauge  and  conformal gauge results}

To check   the equivalence between the  static gauge  and  conformal gauge results
one needs to  verify  the factorization relation \rf{eqi}. 
This  can be  done numerically, as follows. To  evaluate the left hand side of (\ref{eqi}) we used 
  the Gel'fand-Yaglom method (for details, see Appendix B) 
to  compute  numerically the  determinant of the operator $\O_{t\rho\phi}$ in (\ref{Q1})
 as a 
function of  $\Omega$  for various values of k. 
The right hand side of (\ref{eqi}) can be computed directly using 
the expression for the determinant of $\O_\phi$  found above \rf{phj}.
We find  perfect agreement. In Figure 6 we have  
plotted  the expressions on both sides of (\ref{eqi})
 as functions of $\Om$  for $k={1 \ov \sqrt{10}}$. Similar agreement is found for any $k$.

\begin{figure}[hbtp]
\begin{center}
\includegraphics[scale=0.35]{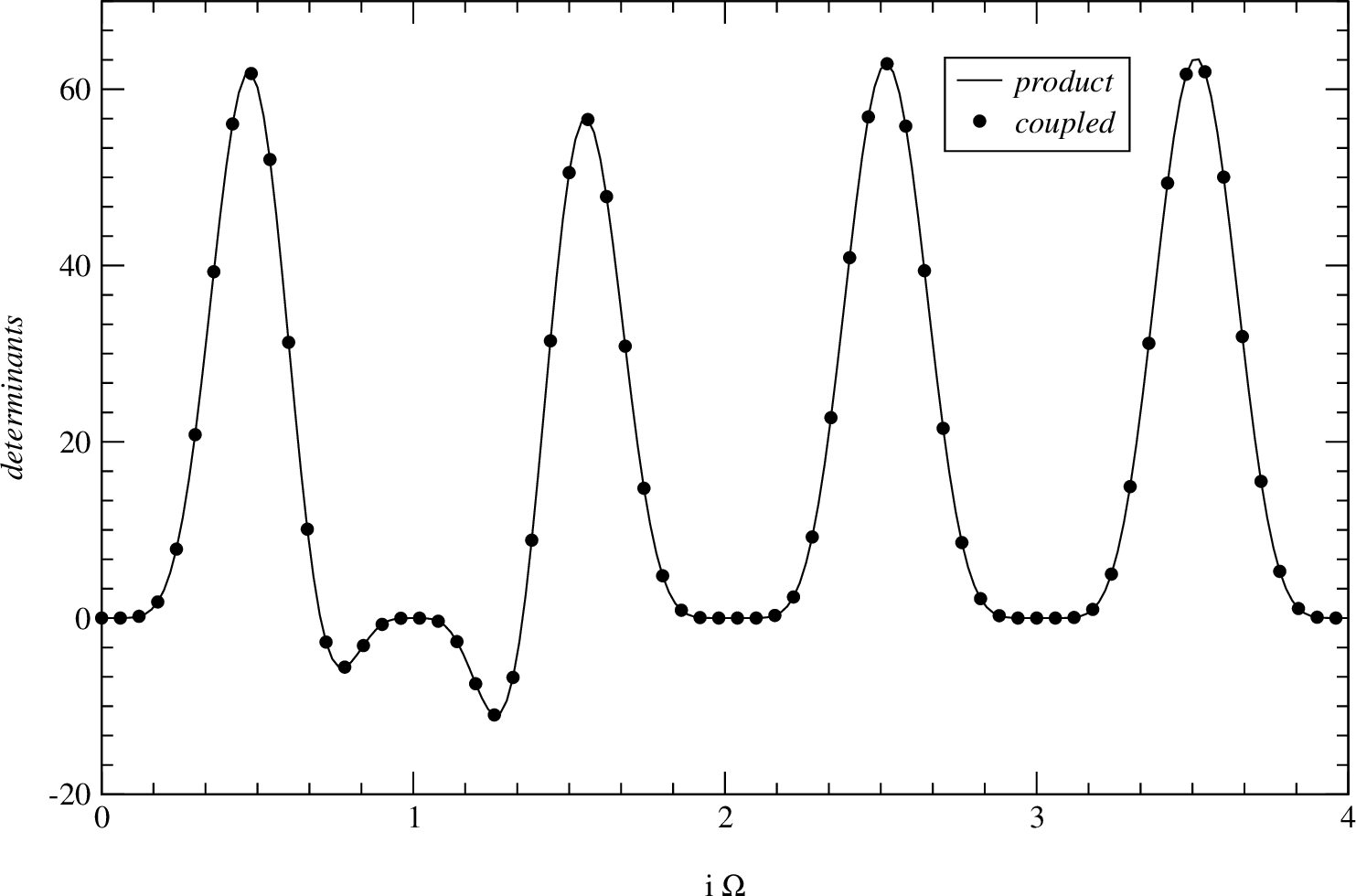}
\caption{Comparison between the left-hand-side  and the right-hand-side
  of Eq.(\ref{eqi}), for  $k={1 \ov \sqrt{10}}$. The circles represent 
  the numerical Gel'fand-Yaglom result  for the determinant $\det \O_{t\r\phi}$ of the three
  coupled fluctuations in
   conformal gauge, while  the solid line is a plot of the corresponding  analytic 
   static gauge expression, given by the product of  the determinant  (\ref{detphi2}) for 
   the massive fluctuation
    $\phi$  and the square  of  the determinant  (\ref{dettrivial}) of a massless mode.  To emphasize the precision of the  agreement we have plotted the oscillatory form, as a function of $i\,\Omega$.}
\end{center}
\label{comparison}
\end{figure}

\subsection{General form of  the 1-loop correction $E_1$ }

Going back to the complete  expression for $E_1$ in \rf{final}
it is useful, in order to safely expand in one of the interesting limits analyzed below,  to separate  there the contributions of  the  massless modes of   $\O_0$ 
(i.e.  $\Om^2$) and 
the  lowest eigenvalues  ($\bar\La_1$ in Table 1) of $\O_\b,\O_\p$  and $\O_{\psi}$.
Then we get 
\begin{equation}
 E_1=-\frac{{1}}{4\pi\k }\int_{-\infty}^{\infty}
 \mathrm{d}\Om\,\left[\ln \frac{(\det'{\cal O}_\psi)^8}{    \det' {\cal O}_\phi\   
 (\det'{\cal O}_{\beta})^2 \   (\det'{\cal O}_0)^5
   } + h(\Om)\right]
\end{equation}
where
\begin{eqnarray}\label{detprime}
{\det}'{\cal O}_{\beta,\p,\psi} &\equiv &\frac{\det{\cal O}_{\beta,\p,\psi}}{\bar\La_1}\,
\quad , \quad \det' {\cal O}_0 \equiv  \frac{\det \O_0}{\Omega^2} \\
 h(\Om)&=&2\ln(\Om^2)-2\ln(\Om^2+\k^2) \ .  \la{hj}
\end{eqnarray}
Using that 
\be\label{des}
\int_{-\infty} ^\infty d\Om\  h(\Om^2)=- 4 \,\pi\,\k
\ee
the one-loop correction to the energy (\ref{final}) takes the form 
\begin{equation}\label{final2}
 E_1=1-\frac{1}{4\pi\kappa}\int_{-\infty}^{\infty}\mathrm{d}\Om \ 
 \,\ln \frac{(\det'{\cal O}_\psi)^8}{\det'{\cal O}_\phi \ (\det'{\cal O}_{\beta})^2 \   (\det'{\cal O}_0)^5} \ . 
\end{equation}
This expression is  straightforward  to evaluate  numerically for various values of 
$k$ or the spin $\S$ in (\ref{Scl}), 
and thus to plot $E_1$.


To gain  more analytic control over the form of $E_1$ as a function of spin $\S$ 
we may consider the expansion of  it in the  large spin (``long string'' or $k\to 1$) limit 
or in the small spin (``short string''  or $k\to 0$) limit.
This will be done in detail  in the following two sections 6 and 7 respectively.
In figure \ref{exactVSnumerical}  we  presented together the results -- 
the plots of $E_1(k) $ found analytically 
in the  large spin expansion (right-most green curve) and  
in the small  spin expansion (left-most red  curve)
and also the plot of the exact $E_1$  found numerically from \rf{final2}
(blue curve connecting the two asymptotic ones). 
As one can see,  already the few leading  terms in the two  respective analytic 
expansions give a very good approximation to  the  exact result. 

\begin{figure}[hbtp]
\begin{center}
\includegraphics[scale=0.6]{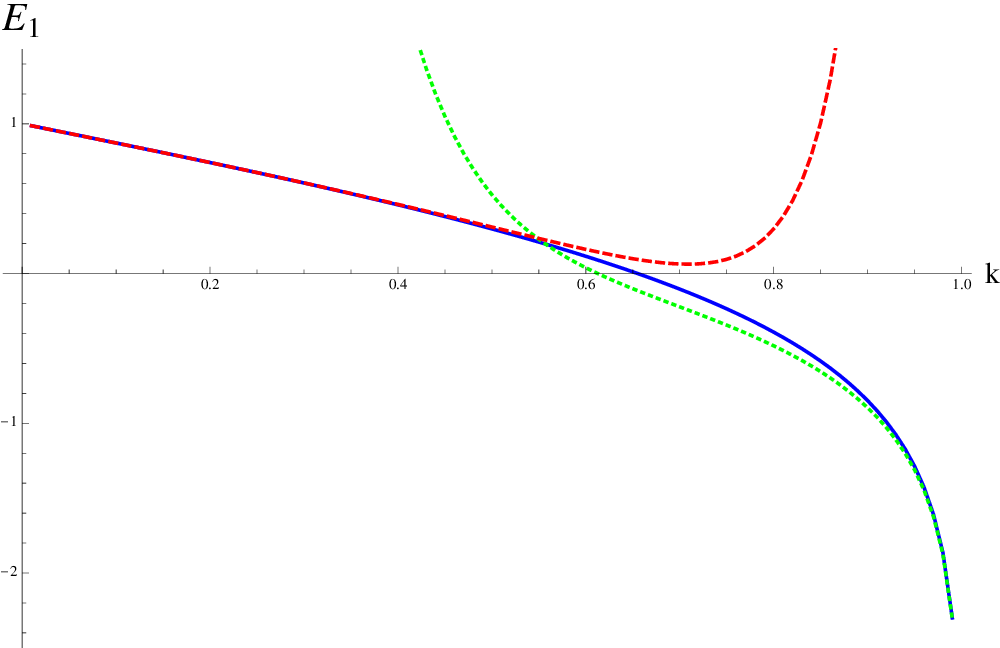}
\end{center}
\small
  \caption{Plots  of $E_1$ as a function of $k$: the blue, solid, curve is found numerically from the exact 
    expression \rf{final2} for generic  values of $k$;
 the green, dotted, curve is found from an analytic expansion in the 
  $k\to 1 $ or large  spin limit, using  the first two terms in (\ref{ordereta3});
   the red, dashed, curve is found from an analytic expansion in the 
  $k\to 0 $ or small spin limit, using  the first two terms in (\ref{final3}). The agreement is excellent in both extreme limits.}
\label{exactVSnumerical}
\end{figure}
\normalsize

\begin{figure}[hbtp]
\begin{center}
\includegraphics[scale=0.7]{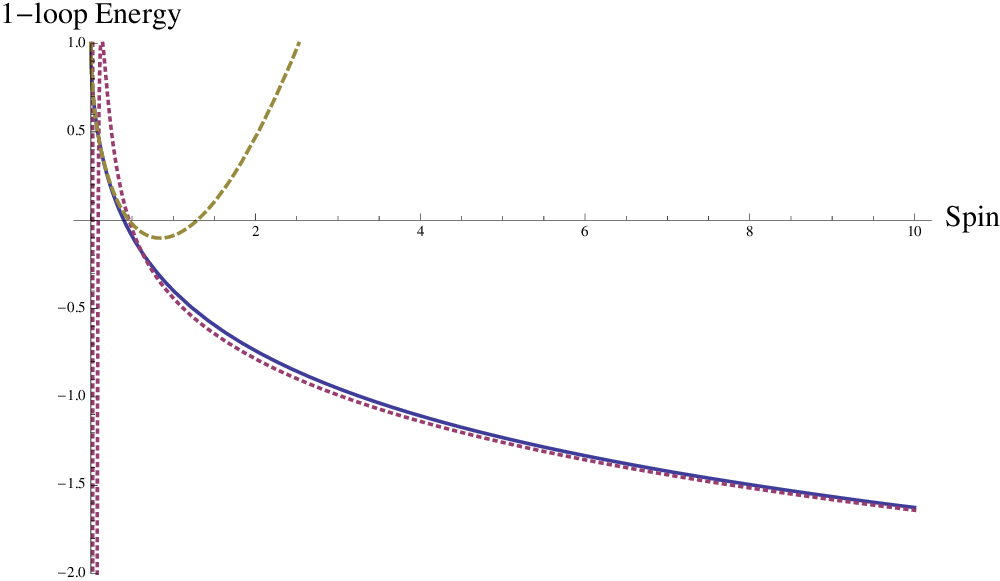}
\end{center}
\small
\caption{Plots  of $E_1$ as a function of the classical spin ${\mathcal S}$. The solid, blue curve is the exact result, compared with the  red (dotted) curve representing the large spin expansion, see (\ref{E1largespin}), and with the gold (dashed) curve representing  the small spin expansion found below in (\ref{finalshort})-(\ref{nan}).  }
\label{exactVSnumerical-spin}
\end{figure}


%


\def \ap {\approx}

\renewcommand{\theequation}{6.\arabic{equation}}
 \setcounter{equation}{0}

 \section{Large spin  expansion }
 \label{sec:large}
 
 This limit (see \rf{etasmall}) is defined as 
 $k\to1$ or,  equivalently,  
 $\eta\to 0$ in \rf{rho0}. 

\subsection{Leading order}
\label{sec:leading}

In this subsection we will compute the leading term in $k\to1$ expansion
of $E_1$ in (\ref{final})  and also comment on first exponential subleading terms.
 
 At  the  leading order the classical solution 
 is approximated by 
  \begin{equation}
  \r' \approx \k_0\ , \ \ \ \ \r''\approx0\ , \ \ \ \ 
 \omega\approx  \kappa \approx \kappa_0\ ,~~~~~~~~\k_0=\frac{1}{\pi}\ln\frac{16}{\eta} \to \infty  \
 . 
\end{equation}
If we  use these  limiting expressions  directly 
in the 
fluctuation operators then we conclude that their 
 potential  terms become constant 
\be\label{leadingops}
\O_{\beta,0} = -\d^2_\s+2\k_0^2+{\Om}^2\ ,~~~~~~~~~\O_{\p,0} =
-\d^2_\s+4\k_0^2+{\Om}^2\  ,~~~~~~~~\O_{\psi_\pm,0}= -\d^2_\s+\k_0^2+{\Om}^2
\ee
and thus we find from \rf{final} \ci{ft1} 
\ba
E_1^{(0)}= \frac{  1}{2\k_0}\sum_{n=-\infty}^\infty\Big[\sqrt{n^2+4\k_0^2} + 2\sqrt{n^2+2\k_0^2}
+5\sqrt{n^2}-8\sqrt{n^2+\k_0^2}\Big] \label{gamma02}
\ee
where we performed the integration over $\Omega$ before commuting the determinants
defined on a unit circle. 
Using the Euler-MacLaurin formula to transform the sum into an
 integral one finds 
\be\label{Eleading}
E_1^{(0)}=\frac{1}{\k_0}\Big[-3\, \k_0^2\ \ln2\    -\frac{5}{12}+\O(e^{-2\pi\k_0})
\Big]\  ,~~~~~~~~~\k_0\to\infty \ , 
\ee
where  the 
leading term is the result of~\cite{ft1} and  the subleading term appeared  in 
\cite{sak}.

Let  us now see  what we get if  we start instead with  
 the exact expressions for  the  determinants (\ref{detbeta2})-(\ref{detpsi2}).
Using the expressions 
 collected in Appendix C (see (\ref{2KZbeta})-(\ref{2KZpsi})), we get  at the leading order 
\ba\label{x}
2\KK(k^2)\,Z(\a_\b\,|\,k^2)&\approx &\pi\,\k_0\,x
\qquad \qquad\quad {\rm with}\quad\quad\quad ~x=\sqrt{2+\frac{\Om^2}{\k_0^2}}\ , \\\label{y} 
2\KK(\td k^2)\,Z(\a_\p\,|\,\td k^2)&\approx &\pi\,\k_0\,y 
\qquad \qquad\quad {\rm with}\quad\quad\quad y=\sqrt{4+\frac{\Om^2}{\k_0^2}}\ , \\\label{z}
\KK(\td k^2)\,Z(\a_\psi\,|\,\td k^2)&\approx &\pi\,\k_0\,z \qquad \qquad\quad {\rm with}
\quad\quad\quad z=\sqrt{1+\frac{\Om^2}{\k_0^2}} \ , 
\ea
which would give, once substituted  into
 (\ref{detbeta2})-(\ref{detpsi2}), the following expressions for the determinants 
\ba\label{detbeta0}
\det\O _{\beta}(\Om^2)&\ap &~~4\sinh^2\left[\pi\,\k_0\,x\right]\\\label{detphi0}
\det\O _{\phi}(\Om^2)&\ap&~~4\sinh^2\left[\pi\,\k_0\,y\right]\\\label{detpsi0}
\det\O_{\psi}(\Om^2)&\ap &-4\cosh^2\left[\pi\,\k_0\,z\right]
\ea
Integrating logarithms of  (\ref{detbeta0})-(\ref{detpsi0}) over $\Omega $ 
one gets the result that may be represented also as 
\be\label{shiftedsum}
E_1\ap \tilde  E_1^{(0)}=\frac{1}{2\k_0} \sum_{n=-\infty}^\infty\Big[\sqrt{n^2+4\k_0^2}+
2\sqrt{n^2+2\k_0^2}+5\sqrt{n^2}-8\sqrt{(n+{\textstyle{\frac{1}{2}}})^2+\k_0^2}\Big] \ . 
\ee
Here the  shift $n\to n+\frac{1}{2}$ in the fermionic contribution
 is due to the $\cosh^2$ instead of $\sinh^2$  form of the 
 determinant (\ref{detpsi0}).\footnote{This shift may be formally  interpreted 
 by saying that fermions have antiperiodic  boundary conditions, 
  so that 
 $\ln\det(-\d_\s^2+\Om^2+\k_0^2)
   =\sum_{n=-\infty}^{+\infty}\,\ln[(n+ {\textstyle{\frac{1}{2}}})^2+\o^2+\k_0^2]$.
 This interpretation  is more of a curiousity  and 
 should not be taken literally as this  expression
 was derived in the large $\kappa_0$ limit where the distinction 
 between the  periodic and antiperiodic fermion  boundary conditions 
  is not actually visible.
 }
While this shift does not affect the result for the two leading terms in 
(\ref{Eleading}), it formally 
  changes the form of the subleading  corrections
  (which should not, however, be trusted in the approximation used to arrive 
  at \rf{shiftedsum}).

Indeed, there  is of course no contradiction as the approximation used to derive  \rf{Eleading}
was supposed to be valid only for the leading term in large $\k_0$ expansion, 
i.e. the expressions for the subleading terms should not be trusted 
a priori.  
Still, let us briefly comment on   the exponential corrections
 to the first  two leading terms in (\ref{Eleading})
 comparing  what  follows from \rf{gamma02}
 to what follows from  \rf{shiftedsum}.  As was found  in \cite{SchaferNameki:2005is} 
 using  $\zeta$-function regularization  of
 the sums in (\ref{gamma02})  
\begin{equation}\label{e0bessel}
 E_1^{(0)}=-3\kappa_0\ln2-\frac{5}{12\kappa_0}-\frac{1}{\pi}\sum_{n=1}
 ^{\infty}\frac{1}{n}\left[K_1(4\pi n\kappa_0)+
 \sqrt{2}K_1(2\sqrt{2}\pi n\kappa_0)-4K_1(2\pi n\kappa_0)\right],
\end{equation}
where $K_1$ is the Bessel function of the second type
\begin{equation}
 \int_{m}^{\infty}\mathrm{d}x\sqrt{x^2-m^2}e^{-2\pi kx}=\frac{m}{2\pi k}K_1(2\pi km).
\end{equation}
The $K_1$  terms represent the exponential corrections since
\begin{equation}\label{bessel}
 K_1(y) \to \sqrt{\frac{\pi}{2y}}e^{-y}\ \big[1+\mathcal{O} ({y}^{-1})\big]\ ,\qquad\ \  y\to\infty.
\end{equation}
Repeating  the same  computation  in the case of (\ref{shiftedsum})
one finds that  
\begin{equation}\label{e0bessel1}
\tilde  E_1^{(0)}=   E_1^{(0)} + \frac{4}{\pi}\sum_{n=1}^{\infty} 
[ (-1)^n - 1] K_1(2\pi n\kappa_0) \ .
\end{equation}



\subsection{Beyond the  leading order}

To find  subleading corrections in large $\k$ 
let us  add and subtract the leading order
 contribution (\ref{Eleading}) from the expression  (\ref{final}):
\begin{eqnarray}\label{gamm}
 E_1&=& { 1 \ov \k} [ -3\kappa_0^2\,\ln 2-\frac{5}{12} + \O(\eta^2)  ]    +  E^{(sub)}_1 \ , \ \ \ \ \ \ \ 
 ~~~~~~~~~~~~\k_0\to\infty\ \\
 E^{(sub)}_1& =& 
 -\frac{\k_0}{4\pi\k } \,\int^\infty_{-\infty}
 \mathrm{d}\bar\Om\,\ln\frac{{\cal D}_\psi^8}{{\cal D}^2_\b\,{
 \cal D}_\p}.
 \end{eqnarray}
Here  we defined 
\be\label{detsreferred}
{\cal D}_\b= \frac{\det\mathcal{O}_{\beta}}{\det\mathcal{O}_{\beta,0}},~~~~~
{\cal D}_\p= \frac{\det\mathcal{ O}_{\phi}}{\det\mathcal{O}_{\p,0}},~~~~~~
{\cal D}_\psi=\frac{\det\mathcal{ O}_{\psi}}{\det\mathcal{ O}_{\psi,0}}
\ee
and introduced $\bar\Om={\Om\over \k_0}$  which is the argument  the integrand 
according to 
 (\ref{x})-(\ref{z}).
  
Expanding the arguments of the determinants, one finds (see (\ref{2KZbeta})-(\ref{2KZpsi}))
\ba
2\KK\,Z(\a_\b\,|\,k^2)&\ap&\pi\,\k_0\,x -2\tanh^{-1}x \\  
2\widetilde{\KK} \,Z(\a_\p\,|\,\td k^2)&\ap&\pi\,\k_0\,y -2\,\tanh^{-1}\frac{y}{2} \\ 
\widetilde{\KK}\,Z(\a_\psi\,|\,\td k^2)&\ap&\pi\,\k_0\,z -\tanh^{-1}z  
\ea
and therefore
\ba\label{ratiocoth1}
\D_\b&=& \frac{\sinh^2[2\KK\,Z(\a_\b\,|\,k^2)]}{\sinh^2[\pi\,\k_0 \,x]}
\ \ap \ \Big[\frac{x^2+1}{x^2-1}-\frac{2x}{x^2-1}\,\coth(\pi\,\kappa_0\,x)\Big]^2\\\label{ratiocoth2}
\D_\p &=& \frac{\sinh^2[2\widetilde{\KK}\,Z(\a_\p\,|\,\td{k}^2)]}{\sinh^2[\pi\,\k_0 \,y]}
\ \ap \ \Big[\frac{y^2+4}{y^2-4}-\frac{4\,y}{y^2-4}\,\coth(\pi\,\kappa_0\,y)\Big]^2\\\label{ratiocoth3}
\D_\psi &=&\frac{\cosh^2[\widetilde{\KK}\,Z(\a_\psi\,|\,\td{k}^2)]}{\cosh^2[\pi\,\k_0 \,z]}
\ \ap \ \frac{1}{1-z^2}\big[1-z\,\tanh(\pi\,\k_0 \,z)\big]^2
\ea
Neglecting the  $\tanh$ and $\coth$ terms in the square brackets 
for large $\kappa_0$ one finds that the second 
contribution in (\ref{gamm}), first contribution at next-to-leading order,
 results in
\ba\no
 E^{(sub)}_1 &\approx &
-\frac{\k_0}{4\pi\k }\,\int_{-\infty}^{+\infty}\mathrm{d}\bar\Om\ \ln
\Big[\Big(\frac{1-\sqrt{1+\bar\Om^2}}{1+\sqrt{1+\bar\Om^2}}\Big)^8\,\Big(\frac{1-
\sqrt{2+\bar\Om^2}}{1+\sqrt{2+\bar\Om^2}}\Big)^{-4} \,\Big(\frac{2-\sqrt{4+\bar\Om^2}}{2
+\sqrt{4+\bar\Om^2}}\Big)^{-1}\Big]\\\label{gammaNLO}
&=& 1 + { 6\ov \pi} \,\ln 2  \  ,~~~~~~~~~~~~~~~~~\k_0\to\infty \ , 
\ea
where we set $\k \ap \k_0$. 
The same result for  this subleading coefficient was found 
in \ci{gr3} using the integrability (algebraic curve) approach (see also \ci{bftt}  and \ci{fz}). 
As discussed in~\cite{bftt} this correction 
 should be due to the near turning point contribution
  that is lost in the naive  approach that 
  treats the potential terms  perturbatively.


Proceeding to the next  order  $\sim \eta= k^{-2} -1 $, 
the evaluation of the various functional determinant ratios gives 
\ba
\D_\b &=& \frac{(x-1)^2}{(x+1)^2}\,\left[1+\frac{2}{x}\,\eta -\frac{\eta}{\pi\,\kappa_0}\,
\frac{2\,(x^2-2)}{x\,(x^2-1)}
+{\cal O}(\eta^2)\right],\\
\D_\p  &=& \frac{(y-2)^2}{(y+2)^2}\,\left[1+\frac{4}{y}\,\eta-\frac{\eta}{\pi\,\kappa_0}\,
\frac{4}{y}
+{\cal O}(\eta^2)\right], \\
\D_\psi &=& \frac{1-z}{1+z}\,\left[1+\frac{1}{z}\,\eta-\frac{\eta}{\pi\,\kappa_0}\,\frac{1}{z} + 
{\cal O}(\eta^2)\right] \ , 
\ea
so that 
\be
\ln\frac{{\cal D}_\psi^8}{{\cal D}^2_\b\,{\cal D}_\p} =e(\bar\Om) +\eta\,\Big[ f(\bar\Om)
+\frac{g(\bar\Om)}{\pi\,\kappa_0}  \Big]+ ... \ ,
\ee
where $e(\bar\Om)$ is the integrand in (\ref{gammaNLO}).
The functions $f(\bar\Om)$ and
 $g(\bar\Om)$ are 
\ba
f(\bar\Om) &=&   \frac{8}{\sqrt{1+\bar\Om^2}}-\frac{4}{\sqrt{2+\bar\Om^2}}-\frac{4}{\sqrt
{4+\bar\Om^2}}\\
g(\bar\Om) &=& - \frac{8}{\sqrt{1+\bar\Om^2}}+\frac{4\,\bar\Om^2}{(1+\bar\Om^2)\,\sqrt{2+
\bar\Om^2}}+\frac{4}{\sqrt{4+\bar\Om^2}}
\ea
and their integrals take the values
\be
\int_{-\infty}^{+\infty} d\bar\Om\ f(\bar\Om) = 12\,\ln\,2\ ,~~~~~~~~~~~~~~
\int_{-\infty}^{+\infty} d\bar\Om\ g(\bar\Om)  = -2\,(\pi+6\,\ln\,2)\ .
\ee
We conclude that to order $\eta$ the large $\k$  expansion 
of the 1-loop energy reads
\bea
&&E_1=\frac{1}{\k}\Big[-3\, \k_0^2\ \ln2\ -\frac{5}{12}+\big(1+\frac{6}{\pi}\,\ln 2 \big)\,\k_0 \no \\
&& \ \ \ \ \ \ \ \ \ \ \ \ \ \ 
-\ \frac{1}{\pi} \big[3\, \k_0\, \ln 2\ -\frac{1}{2}\,\big(1+\frac{6}{\pi}\,\ln\,2 \big)\big]\ \eta 
 + \O(\eta^2) \Big]\  ,  \ \ \ \ \ \ \ \ \ \k_0\to\infty \ . \la{resoi}
\eea
Here we did not expanded explicitly the overall factor of $ 1 \ov \k$, 
\be \la{kep}
{1 \ov \k} ={1 \ov \k_0}  \Big[1  +  {1 \ov 4} ( 1 - {2 \ov \pi\k_0})  \eta + \O(\eta^2) \Big] \ , \ \ \ \
\ \ \ 
\eta= 16 e^{-\pi \k_0} \to 0  \ .
\ee  
The coefficient of the leading $\eta $ correction is in agreement with the one found in 
\cite{bftt}, while the next $ { \eta \ov \k_0}$  term is a new result.

In going  to higher than first orders of expansion in $\eta$ there is a potential problem of accounting for
the contributions of terms like $\coth (\pi\k_0 x) -1$ in \rf{ratiocoth1}- (\ref{ratiocoth3})
we have dropped above. For example, 
\be\label{expcorr}
 e^{-2 \k_0 \pi  z} \sim\,\Big(\frac{\eta}{16}\Big)^2\Big[1-\frac{\pi \Om^2}{ \k_0}+
\frac{\pi ^2 \Om ^4}{2 \k_0^2}+\frac{3  \pi ^4  \Om^4-2 \pi ^6 \Om ^6}{12  \k_0^3 \pi ^3}+...\Big]\ ,
\ee
and similar terms arise also in the expansion of the reference determinants, see (\ref{bessel}).
 Such terms need to be resummed. 
 and, while there is the possibility that all such terms may cancel, this is   not clear at the moment. 
 In Appendix D we present the evaluation of the leading large $\k_0$ correction to the one-loop energy due to these contributions, 
 while in Appendix E we consider a different type of expansion in the $k\to 1$ limit.

Ignoring this complication,  we  have found   that 
the one-loop energy  has the  following  structure of large spin 
 expansion 
\ba\no
E_1 &=& \frac{\kappa_0}{\kappa} \,\Big[\big(c_{01}\,\kappa_0 + c_{00} +
\frac{c_{_{0,-1}}}{\kappa_0}\big)+ \,\big(c_{11}\,\k_0+c_{10} + 
\frac{c_{_{1,-1}}}{\kappa_0}\big) \eta
+  \\\label{E1large}
& & \ \ \ + \ \big(c_{21}\,\kappa_0 + c_{20} + 
\frac{c_{_{2,-1}}}{\kappa_0}\big)\eta^2 + 
 \,
\big(c_{31}\,\kappa_0 + c_{30} + \frac{c_{_{3,-1}}}{\kappa_0}\big)\eta^3 +\O(\eta^4) \Big] \ ,
\label{ordereta3}\ea
where the explicit values are 
\ba\label{coefficientsfirst}
c_{01} &=& -3   \ln 2\ , ~~~~~~~~~~~~
c_{00} = 1 +{6\ov \pi} \ln 2 \, ~~~~~~~~~~~~
c_{_{0,-1}} = -\frac{5}{12}\ , \\
c_{11} &=& 0 \ ,~~~~~~~~~~~~~~~~~~~~~
c_{10} = -{3\ov \pi} \ln 2 \,~~~~~~~~~~~~~~~
c_{_{1,-1}}=\frac{1}{2\pi }+\frac{3\,\ln2}{\pi^2}\ ,  \\
c_{21} &=& -\frac{\pi }{32}-\frac{3}{32}   \ln 2\ ,~~~~~~~
c_{20} = \frac{1 }{16}+\frac{39 \ln 2}{32\pi}\ ,~~~~~~~
c_{_{2,-1}} = -\frac{13}{64\pi}-\frac{63 \ln 2}{32\pi^2 }\ ,\\\label{coefficientslast}
c_{31} &=& \frac{\pi }{32}+\frac{3 }{32}  \ln 2\ ,~~~~~~~~~~
c_{30} = -\frac{3  }{32}-\frac{13 \ln 2}{16\pi }\ ,~~~~~~
c_{_{3,-1}} = \frac{29}{192\pi}+\frac{85 \ln 2}{64 \pi^2 }\ .
\ea
For completeness, we report here the first few orders in the large spin expansion of the 1-loop energy as found using (\ref{etasmall}) in (\ref{E1large})
\ba\no
&&\!\!\!\!\!\!\!\!\!\!E_1= -\frac{3 \ln2}{\pi } \ln\bar{\S}+\frac{\pi+6 \ln2}{\pi
   }-\frac{5 \pi }{12 \ln \bar{\S}}
   -\frac{1}{\bar\S}\Big[\frac{24 \ln2}{\pi } \ln \bar{\S}-\frac{4\pi+36 \ln2}{\pi
   }+\frac{5 \pi }{3 \ln^2\bar\S }\Big]+\O\left(\frac{1}{\bar{\S}^2}\right)\\\label{E1largespin}
 &&\!\!\!\!\!\!\!\!\!\!  \bar{\S}=8\,\pi\,\S,~~~~~~\S\gg1
\ea

\subsection{Test of reciprocity}

With the expressions 
 (\ref{ordereta3})-(\ref{coefficientslast}) at hand, we are able to
 the confirm  and 
extend the analysis of~\cite{bftt}, in which the reciprocity relations
between the coefficients in large spin expansion of the energy (or twist 2 anomalous 
dimension at strong
coupling)  \cite{bk,dok2,Beccaria:2009vt}  were  checked up to order $\eta$.

To do this  one needs to determine the functions\foot{ 
${\cal E}_0$ and ${\cal E}_1$ are the classical and the  1-loop energies rescaled by a factor string
tension.}
\be \Delta({\cal S}) = \Delta_0 + { 1 \ov \sql}  \Delta_1 + ... \ , \ \ \ \ \ \  
 \Delta_0  = {\cal E}_0({\cal S})-{\cal S}, \qquad \Delta_1 = {\cal E}_1({\cal S}),
\ee
as functions of the spin ${\cal S}$, which as at the classical level is obtained by 
replacing the parameter $\eta$ with its expansion   $\eta = \eta({\cal S})$ in terms 
of the spin (\ref{etalarge}).
One is then to compute the function ${\cal P}$ defined by 
\ba\label{defP}
\Delta({\cal S})  &=&  {\cal P}({\cal S} + \frac{1}{2}\,\Delta({\cal S}))   \ .  
\ea
The test of reciprocity amounts   to  the check of parity of ${\cal P}({\cal S})$ under 
${\cal S}\to -{\cal S}$.
Solving the functional  equation  in (\ref{defP})  as 
\be
{\cal P}({\cal S}) = \sum_{k=1}^\infty\frac{1}{k!}\left(-\frac{1}{2}\frac{d}{d{\cal S}}
\right)^{k-1}\big[\Delta({\cal S})\big]^k \ , 
\ee
and expanding  the function ${\cal P}$  in $1\ov \sqrt\lambda$
\be
{\cal P} = {\cal P}_0 + \frac{1}{\sqrt\lambda}\,{\cal P}_1 + \cdots,
\ee
one finds
\ba
{\cal P}_0({\cal S}) = \sum_{k=1}^\infty\frac{1}{k!}\left(-\frac{1}{2}\frac{d}{d{\cal S}}
\right)^{k-1}[\Delta_0({\cal S})]^k, \\
{\cal P}_1({\cal S}) = \sum_{k=1}^\infty\frac{1}{k!}\left(-\frac{1}{2}\frac{d}{d{\cal S}}
\right)^{k-1}[k\,\Delta_0({\cal S})^{k-1}\,\Delta_1({\cal S})].
\ea
Working out ${\cal P}_1$ and looking at all terms which are odd under ${\cal S}\to 
-{\cal S}$ we find that they vanish 
if the following reciprocity constraints hold
\ba
c_{10} &=& \frac{1}{\pi}\,c_{01}\ , \ \ \ \ \ \ \ \
c_{_{1,-1}} = \frac{1}{2\pi}\,c_{00}\ , \ \ \ \ \ \ \ 
c_{31} = -c_{21}\ , \\
c_{30} &=&-c_{20}-\frac{1}{6\pi} c_{01}+\frac{1}{\pi}\,c_{21}\ ,  \\
c_{_{3,-1}} &=& -c_{_{2,-1}}+\frac{1}{4\pi^2}\,c_{01}-\frac{1}{12\pi} c_{00}+\frac{1}{2\pi}\,c_{20}.
\ea
As usual, the coefficients of terms with odd powers of $\eta= {2 \ov \S} + ...$  in \rf{etasmall}
are determined  by coefficients of  terms with even powers 
of $\eta$.
 Using the list of 
explicit coefficients found above (\ref{coefficientsfirst})-(\ref{coefficientslast}),
 we find 
that these relations are indeed satisfied.

\renewcommand{\theequation}{7.\arabic{equation}}
 \setcounter{equation}{0}
 
\section{Small spin  expansion }

The small spin or short string limit~\cite{tt,rt} is realized
by  sending $\eta\to\infty$ or $k\to 0$ (see section 2). 

The general expansion of the determinants, see (\ref{derZ})-(\ref{detbetashort}),  has the form 
\be
\det\,{\cal O}_{f  } = D_f^{(0)}(\Om)  + \frac{1}{\eta}\,D_f^{(1)}(\Om)  
+ \frac{1}{\eta^2}\,D_f^{(2)}(\Om)  + \cdots, \ \ \ \ 
f= (\beta, \phi, \psi)  \ , 
\ee
where
\bea
&&D_\beta^{(0)}(\Om) = D_\phi^{(0)}(\Om) = D_\psi^{(0)}(\Om) = 4\,\sinh^2(\pi\,\Om), \la{0}\\
&& D_\beta^{(1)}(\Om) =\frac{2 \pi  \sinh (2 \pi  \Om )}{\Om }, ~~~
D_\phi^{(1)}(\Om) = \frac{4 \pi\Om  \sinh (2 \pi  \Om )}{\Om ^2+1}, ~~~
D_\psi^{(1)}(\Om) = \frac{4 \pi  \Om \sinh (2 \pi  \Om)}{4 \Om^2+1}  , 
\no \\
&&
D_\beta^{(2)}(\Om) =\frac{\pi ^2 \cosh (2 \pi \Om )}{\Om ^2}-\frac{\pi  \left(3 \Om ^4
+6 \Om ^2+2\right) \sinh (2 \pi  \Om )}{4 \left(\Om ^5+\Om ^3\right)}\no \\
&& D_\phi^{(2)}(\Om) = \frac{4 \pi ^2 \Om ^2 \cosh (2 \pi \Om )}{\left(\Om ^2+1\right)^2}
-\frac{\pi  \Om  \left(3 \left(\Om ^2+4\right) \Om ^2+1\right) \sinh (2 \pi  \Om )}{2
 \left(\Om^2+1\right)^3}, \no  \\
&& D_\psi^{(2)}(\Om) = -\frac{\pi  \Om  \left(48 \left(\Om ^4+\Om ^2\right)+1\right) 
\sinh (2 \pi  \Om )}{2 
\left(4 \Om ^2+1\right)^3}+\frac{4 \pi ^2 \Om ^2 \cosh (2 \pi  \Om )}{\left(4 \Om^2+1\right)^2}. \la{ded}
\ea
The first correction to the quantity entering the effective action (\ref{final}) is 
\ba
\label{IR}
\ln \frac{\det^8{\cal O}_\psi}{ \det{\cal O}_\phi\, \det^2{\cal O}_{\beta} \, \det^5{\cal O}_0} =
\frac{1}{\eta}\,\frac{2 \pi  \left(2 \Om^2-1\right)
 \coth (\pi  \Om )}
{\Om  \left(\Om ^2+1\right) \left(4 \Om^2+1\right)}+{\cal O}\left(\frac{1}{\eta^2}\right),
\ea
which is integrable at $\Om\to\infty$ but has a pole at   $\Om=0$, 
i.e.  produces an IR divergence. 
Such an IR effect disappears by integrating separately the lowest eigenvalues (see Table 1), which,  in
 fact,   behave as zero modes around $\Om\sim0$ (in the case of the $\beta$ 
 fluctuation this only happens in the short string limit $\eta\to\infty$)
\be
\bar\Lambda^{(\b)}_1 = \Om^2 + \frac{1}{\eta} + \cdots,~~~~~~~~~~~~~
\bar\Lambda^{(\p)}_1 = \Om^2 ,~~~~~~~~~~~~~~
\bar\Lambda^{(\psi)}_1 = \Om^2 .
\ee
This is equivalent to use the definition  (\ref{final2}) for the 1-loop correction to the 
energy.  Indeed, with the definition (\ref{detprime})
 the quantity one is to evaluate 
\ba
\ln \frac{(\det'{\cal O}_\psi)^8}{\det'{\cal O}_\phi\, (\det'{\cal O}_{\beta})^2\,  (\det'{\cal O}_{0})^5} =
\frac{1}{\eta}\,
\frac{2 \left[4 \Om ^4+5 \Om ^2+\pi  \left(2 \Om ^2-1\right) \Om  \coth (\pi  \Om )+1\right]}
{ \Om ^2 \left(4 \Om ^4+5 \Om ^2+1\right)}+...
\ea
is now finite and can be integrated to give $2\pi(8\ln 2-3)$.
On the other hand, the contribution of the lowest eigenvalues has been shown to give a finite number at (\ref{des}) at all orders in $1/\eta$.

Going to one further order in the large $\eta$ 
expansions of the determinants 
and adding  all together  one finds 
for the expansion of the 1-loop energy (\ref{final2})
\ba\no
 E_1 &=&1-\frac{1}{4\pi\kappa}\int_{-\infty}^{\infty}\mathrm{d}\Om \ \ln \frac{(\det'{\cal O}_\psi)^8}{(\det'{\cal O}_{\beta})^2 \det'{\cal O}_\phi\,(\det'{\cal O}_0)^5}\\\no
 &=&1+ \frac{1}{\k} \Big[
 \textstyle{\big(\frac{3}{2}-4\,\ln2\big)}\eta^{-1} -
\textstyle{\big(1-\frac{3}{2}\,\ln\,2-\frac{3}{8}\,\zeta(3) \big)}\eta^{-2} \\
\label{final3}
  &&\ \ -\ \textstyle{\big(
-\frac{27}{16}+\frac{7}{4}\,\ln\,2+\frac{9}{32}\,\zeta(3) +\frac{15}{32}\,\zeta(5)
\big)}\eta^{-3}  + \O (\eta^4) \Big] \ .
\ea
Here we did not  expand explicitly  the factor 
\be \la{kah}
{1 \ov \kappa} =  \sqrt{ \eta} \, \big[1 + {1 \ov 4  }\eta^{-1}  +   \O(\eta^{-2}) \big] \ . 
\ee
Substituting the expansion of $\eta$ in terms of the spin (\ref{etalarge}), we can finally obtain the following small spin expansion of the 1-loop correction
to the energy 
\ba\la{su}  E_1 &=& E_1^{(\rm an)} + E_1^{(\rm nan)} \ , \\ 
E_1^{(\rm an)}&=&\sqrt{2\,\S}\Big( \textstyle{[\frac{3}{2}-4 \ln\,2]+[-\frac{23}{16}+\frac{3}{2} \ln\, 2+\frac{3}{4}\, \zeta(3)]\,\S }  \\\label{finalshort}
&&+ \textstyle{ [\frac{689}{256}-\frac{63}{32}\, \ln\, 2-\frac{15}{32}\, \zeta (3)-\frac{15}{16}\, \zeta (5)]}\,\S^{2}+\O(\S^{3})\Big) \ , \\
\label{nan}
E_1^{(\rm nan)}&=&1+\O(\S),\quad\quad\quad\quad\quad\quad\quad\quad\quad\quad\quad\quad\quad\quad\quad\quad\quad\quad\quad\quad\quad \ . 
\ea
We have separated $E_1$, as in ~\cite{rt}, into 
 an ``analytic'' part (with $\S$-dependence  similar to the 
 classical energy (\ref{Ec})) and  a ``non-analytic'' part, containing 
 ``would-be IR singular''  contributions of the   lowest eigenvalues.

We conclude that the procedure adopted in this paper leads to the
 same structure of the small spin  expansion of
the 1-loop energy as found   in ~\cite{tt}.
The coefficients of the transcendental terms proportional to 
   $\ln2$ and  $\zeta(3)$ in \rf{finalshort} are exactly the same as in 
   \ci{tt} (see eq. (4.37) there). 
The coefficients of the rational terms are, however, different. 

Let us note that  a
 separate   treatment of the zero-mode contribution in~\cite{BTunpublished}
  led also to  a different result for (\ref{nan}) (cited in eq. (3.60) in~\cite{rt}).
 Refs. \cite{tt,BTunpublished} used  the  standard ``near-flat-space'' perturbation theory treatment of the determinants in the conformal gauge. 
 A  disagreement  with our present results
 is apparently due to the prescription  adopted in~\cite{tt,BTunpublished} for the  
 projecting out the zero mode contributions.
  As discussed  in Appendix F, a somewhat different prescription would lead to the same result  as the  one that
  one obtains  using the small spin  perturbation theory for the  static gauge
  determinants.




\section{Conclusions}

In this paper  we have found an exact expression for the one-loop 
correction to the energy of  the  folded  string  spinning in the $AdS_3$ part of $AdS_5
\times S^5$. The main technical advance is that we have shown
 that all the
fluctuation operators, in the static gauge, have the single-gap Lam\'e
form. As a result,  their  determinants  can be computed in a closed form. We have 
verified explicitly that, as expected, 
 the one-loop energy correction is the same in the static
and the conformal gauges, even though the  structure of the two fluctuation determinant ratios 
 appears to be  quite different. 
 
The analytic
expressions  for the fluctuation determinants 
permitted us to carry out  improved expansions in the small and large spin limit;
 the latter allowed us to verify  that the reciprocity relations continue to be
satisfied at strong coupling.  
Perhaps more importantly, our demonstration that the
semi-classical fluctuation problem is governed by simple finite-gap
operators  gives a new perspective on the role  of integrable systems in
the analysis of quantum corrections in such string models. 
In fact, finite-gap fluctuation
operators are naturally described in terms of algebraic curves of Riemann
surfaces associated with the finite-gap spectrum~\ci{belokos}, making 
the connection with
the classical integrability  (algebraic curve) approach of ~\cite{kaz,gr1} explicit.

The integrability  approach to semi-classical quantization  
relies 
on the classical integrable structure of the theory.
The investigation of  the monodromy of the Lax connection \ci{bpr} 
for the $AdS_5\times S^5$
 superstring action  leads  to the derivation of a spectral curve 
for any solution of the classical string equations of motion~\ci{kaz,BeisertKazakov}. 
This is an example of the general finite gap description of classically integrable theories~\ci{novikov}  which, reformulated in terms of  a Riemann-Hilbert problem, 
leads to certain  integral equations for each finite-gap curve associated to a classical solution. 
The same finite-gap  integral equations happen to  appear in 
 the continuum limit of the (discrete) algebraic Bethe Ansatz  equations \ci{AFS}. 
 
Starting with the  classical  algebraic curve describing a particular solution 
 one can 
develop a  semiclassical quantization  \ci{gr1,vicedo}  by deforming the cuts definining the 
algebraic curve (adding  extra roots) 
~\cite{GrVi1,GrSchaVi}. Fluctuations are then
 perturbations of the cuts, and the one-loop correction to the energy is given as usual by the sum of the energy shifts (or characteristic frequencies) due to these fluctuations. 
 Alternatively, one may try to guess the quantum 
 extension of the classical finite gap integral equations, having as guiding principle the gauge theory information implying a description in terms of an asymptotic    Bethe Ansatz \ci{AFS}. 
Improved by the   phase \cite{BeisTsey,HL}  extracted from the 1-loop string data of~\cite{ptt},  the  Bethe Ansatz result for the 1-loop correction to string energy 
  was shown \ci{GrVi2}  to agree, for a  generic classical superstring solution, 
with the approach based on extracting the characteristic frequencies  by perturbing  the 
algebraic curve.   
 This  general equivalence  was  recently extended to include also 
 the  exponentially suppressed finite size effects  with  the asymptotic Bethe Ansatz 
 starting point replaced by an appropriate Thermodynamic Bethe Ansatz (see ~\ci{gr2}   and references therein).

Comparing this integrability approach 
 to the one of    the present  paper, 
 notice that even if we did not explicitly  refer to the classical integrability of the 
  string sigma model,  
we ``rediscovered'' 
 the integrability at the  one-loop level 
 via the connection with the integrable, finite-gap, Lam\'e equation.

  In addition to  stimulating the study of 
   detailed  relation  between the two 
    approaches 
  at the 1-loop level \cite{bdfgpt}, the   findings of the present papers 
  have a  methodological merit 
  of  explicitly illustrating   on a rather important and non-trivial example of 
    how  the integrability  of the \adss superstring sigma model is extended
     from the classical to
the semiclassical one-loop level.

 This  connection is, of course, not surprising from a general perspective:
  given a set of integrable 
 classical equations,  the  linear problem  for small fluctuations near a given 
  solution  is found  by considering a small variation  of the original non-linear 
  equations and should  thus 
  be essentially   controlled  by the original classical  integrable structure. 
 However, the technical details  of such connection may be quite intricate. 
 The small  fluctuation problem  is, in general,  described  by a complicated 
  coupled set of 
 linear  differential equations, i.e. by a matrix differential operator, 
 while the standard examples of integrable   spectral problems  involve 
 2-nd order ordinary differential operators with their  integrability 
  related to a special  type of their potential terms. 
 The  general study of which kind of integrable matrix  differential operator spectral 
 problems are associated to non-linear string  sigma model  type  classical equations 
 appears to be  an interesting open problem.

The extension of our present results to the case of  spinning folded string with 
a non-zero angular momentum in $S^5$ is currently under investigation~\cite{bdfgpt}.

%

\section*{Acknowledgments }
We thank I. Adam,  T. McLaughlin, C. Meneghelli, R. Roiban,  D. Seminara, S. Theisen,
 A. Tirziu and expecially N. Gromov for  useful discussions.  
 We  are grateful to  R. Roiban for helpful comments on the draft.
 V. F. would like to thank the  Theory Groups of IPhT CEA-Saclay and DESY-Hamburg  for discussions and kind hospitality while parts of this work were done. 
 G. D. is supported in part by  the DOE grant DE-FG02-92ER40716.
V. F. is supported by the Alexander von Humboldt foundation.
 
 \bigskip
 \bigskip
 
 \appendix
 
\section*{Appendix A:  Relevant elliptic function properties and identities }
\refstepcounter{section}
\def\theequation{A.\arabic{equation}}
\setcounter{equation}{0}

\subsection*{Complete elliptic integrals}

The \emph{complete elliptic integrals of the first and second kind} are defined as functions of their modulus $k^2$ as follows
\be
\KK(k^2)=\KK=\int_0^{\frac{\pi}{2}}d\theta\ (1-k^2\,\sin^2\theta)^{-1/2}\, , ~~
~~~~~~~\EE(k^2)=\EE=\int_0^{\frac{\pi}{2}}d\theta \ (1-k^2\,\sin^2\theta)^{1/2}\,
\ee
One also defines the complementary modulus
\be
k'^2=1-k^2~~~~~~~{\rm and}~~~~~~~\KK'(k^2)=\KK'=\KK(1-k^2)=\int_0^{\frac{\pi}{2}} \, d\theta\ 
(1-k'^2\,\sin^2\theta)^{-1/2}\ .
\ee

\subsection*{Jacobi elliptic functions}

Defining the \emph{Jacobi amplitude} as
\be
\varphi={\rm am} (u\,|\,k^2),~~~~{\rm where}~~~~u=\int_0^\varphi d\theta\  (1-k'^2\,\sin^2\theta)^{-1/2}
\ee
the \emph{Jacobi elliptic functions} $\sn,\cn,\dn$  defined by
\be
\sn(u\,|\,k^2)=\sin \varphi,~~~~~~~\cn(u\,|\,k^2)=\cos \varphi,~~~~~~\dn(u\,|\,k^2)=(1-k^2\,\sin^2 \varphi)^{1/2}
\ee
are doubly periodic functions of $u$, with real-valued periods that are either $2\KK$ ($\dn$) or $4\KK$ ($\sn$ and $\cn$) and purely imaginary periods that are either $2i\,\KK'$ ($\sn$) or $4i\,\KK'$ ($\cn$ and $\dn$). The fundamental period-parallelogram for the Jacobi elliptic functions is, therefore, the rectangle with corners at $(0,4\KK,4i\KK',4\KK+4i\KK')$, where zero occur for real values of $u$ (at $2\KK$ and $4\KK$) while singularities occur for imaginary values of $u$ (at $i\,\KK'$ and $3i\KK'$).

Other Jacobian elliptic functions useful for us are
\be
&&{\rm cd}(u\,|\,k^2)=\frac{\cn(u\,|\,k^2)}{\dn(u\,|\,k^2)},~~~~~~~~~~ {\rm sd}(u\,|\,k^2)=\frac{\sn(u\,|\,k^2)}{\dn(u\,|\,k^2)}\\
&& {\rm ns}(u\,|\,k^2)=\frac{1}{\sn(u\,|\,k^2)},~~~~~~~~~~ {\rm nd}(u\,|\,k^2)=\frac{1}{\dn(u\,|\,k^2)}
\ee 

Useful relations between the squares of the functions are
\ba
&&-\dn^2(u\,|\,k^2)+k'^2=-k^2\,\cn^2(u\,|\,k^2)=k^2\,\sn^2(u\,|\,k^2)-k^2\\
&&-k'^2\,{\rm nd}(u\,|\,k^2)+k'^2=-k^2\,k'^2\,{\rm sd}^2(u\,|\,k^2)=k^2\,{\rm cd}(u\,|\,k^2)-k^2.
\ea
A useful representation for $\sn(u\,|\,k^2)$ is
\begin{eqnarray}
{\rm sn}(u\,|\, k^2)=\frac{\pi}{2{\mathbb K}^\prime}\sum_{n=-\infty}^\infty (-1)^n\tanh\left(\frac{\pi}{2{\mathbb K}^\prime}\left( u-2\, n\, {\mathbb K}\right)\right)
\label{largesn}
\end{eqnarray}

\subsection*{Jacobi Eta, Theta and Zeta functions}

The \emph{Jacobi $H$, $\Theta$ and $Z$} functions are defined as follows in terms of the Jacobi $\vartheta$ functions
\be\label{jacobidef}
H(u\,|\,k^2) = \vartheta_1\left(\frac{\pi\,u}{2\,\mathbb{K}}, q\right), \qquad
\Theta(u\,|\,k^2) = \vartheta_4\left(\frac{\pi\,u}{2\,\mathbb{K}}, q\right),\qquad
Z(u\,|\,k^2) = \frac{\pi}{2\,\mathbb{K}}\,\frac{\vartheta_4'(\frac{\pi\,u}{2\,\mathbb{K}}, q)}
{\vartheta_4(\frac{\pi\,u}{2\,\mathbb{K}}, q)}
\ee
where 
\be
q=q (k^2) =  \exp\Big(-\pi\frac{\mathbb{K}'}{\mathbb{K}}\Big).
\ee
Useful periodicities for them are
\ba\label{periodicityH}
H(u+2\KK \,|\,k^2)&=&-H(u \,|\,k^2),\\
\Theta(u+2\KK \,|\,k^2)&=&\Theta(u \,|\,k^2),\\
Z(u+2\KK \,|\,k^2)&=&Z(u\,|\,k^2)
\ea
Useful representations for ${Z}(u\,|\,k^2)$ are the integral representation
\begin{equation}\label{intzeta}
Z(\mathrm{sn}^{-1}(y|k^2)\,|\,k^2)=\int_0^y\mathrm{d}t\left[\sqrt{\frac{1-k^2t^2}{1-t^2}}-\frac{\EE(k^2)}{\KK(k^2)}
 \frac{1}{\sqrt{(1-t^2)(1-k^2t^2)}}\right]
\end{equation}
and
\begin{eqnarray}
{\mathbb Z}(\alpha; k^2)=\int_0^\alpha du \ {\rm dn}^2(u; k^2)  \  - \frac{{\mathbb E}(k^2)}{{\mathbb K}(k^2)}
\, \alpha\ .
\label{zed}
\end{eqnarray}
We also recall the following series representation
\begin{eqnarray}
{Z}(u\,|\, k^2)=-\frac{\pi}{2{\mathbb K}^\prime\,{\mathbb K}}u+\frac{\pi}{2{\mathbb K}^\prime}\sum_{n=-\infty}^\infty \tanh\left(\frac{\pi}{2{\mathbb K}^\prime}\left(  u-2\, n\, {\mathbb K}\right)\right)
\label{largezed}
\end{eqnarray}



\subsection*{Landen transformations useful for folded string fluctuation operators}

\begin{description}

\item \emph{(a) Bosonic fluctuation $\phi$}

Consider the bosonic fluctuation (\ref{ch})-(\ref{phb})
\be\no
\O_{\p} =-\d^2_\s+2\o^2\,k^2\,\sn^2(\o\s+\KK \,|\,k^2)
+2\o^2 \,{\rm ns}^2(\o\s+\KK \,|\,k^2)
\ee
In the potential in $\O_\p$  one can use
 \be\label{phirelations}
   \mathrm{sn}^2(u\,|\,k^2)=\frac{1-\mathrm{cn}(2u\,|\,k^2)}{1+\mathrm{dn}(2u\,|\,k^2)},
 ~~~~~~~~~~\mathrm{ns}(u\,|\,k^2)=i\,\mathrm{sn}(u+i\KK'\,|\,k^2)
 \ee
and perform the  Landen transformation   
 \begin{equation}
   \mathrm{sn}((1+\tilde k')u\,|\,k^2)=(1+\tilde k')\mathrm{sn}(u\,|\,\tilde k^2)\,\mathrm{cd}(u\,|\,\tilde k^2)
  \end{equation}
  with $\td{k}$ defined in (\ref{ktilde}).
Rescaling  the variable
\be\label{rescalephi}
\bar\sigma=\frac{2\omega\sigma}{1+\tilde{k}'}
\ee
one obtains  the single-gap Lam\'e operator 
\ba\label{phibis}
\O_\phi& =&-\d^2_z+2\,\td{k}^2\,\sn^2(z+i\,\widetilde{\KK}'\,|\,\td{k}^2)-\td {k}^2+\frac{\pi^2\,{\Om^2}}{4\,\widetilde{\KK}^2}
\ea
with periodicity $\phi(z)=\phi(z+4\widetilde{\KK})$ in the rescaled variable $z=\frac{2\,\widetilde{\KK}}{\pi}\,\s$. Notice that the imaginary shift of $x$ makes the potential singular,  as it must be from the original form (\ref{phb}), where  singularities  are  manifest at \(\sigma=\left(n+\frac{1}{2}\right)\pi\),
\(n\in\mathbf{N}\). Such imaginary part, however, does not affect the discussion
leading to the determinant expression.~\footnote{In particular, it does not affect the monodromy of the potential.}

\item \emph{(b) Fermionic fluctuations}

Consider the fermionic fluctuations (\ref{ch})-(\ref{pss})
\begin{equation}\no
 \mathcal{O}_{\psi_\pm}=-\frac{\mathrm{d}^2}{\mathrm{d}\sigma^2}+
 \kappa^2\mathrm{sn}^2(\omega\sigma+\KK\,|\, k^2)\pm\,\kappa\,\omega\,\mathrm{cn}(\omega\sigma+\KK\,|\, k^2)\,
 \mathrm{dn}(\omega\sigma+\KK\,|\, k^2)+\Omega^2
\end{equation}

One can use the Landen transformation  
\begin{eqnarray}\label{landen1}
 \mathrm{sn}((1+\tilde k')u\,|\,k^2)&=&(1+\tilde k')\,\mathrm{sn}(u\,|\,\tilde k^2)\,\mathrm{cd}(u\,|\,\tilde k^2),\\
 \mathrm{cn}((1+\tilde k')u\,|\,k^2)&=&\mathrm{nd}(u\,|\,\tilde k^2)-(1+\tilde k')\mathrm{sn}(u\,|\,\tilde k^2)\mathrm{sd}(u\,|\,\tilde k^2),\\
 \mathrm{dn}((1+\tilde k')u\,|\,k^2)&=&\mathrm{nd}(u\,|\,\tilde k^2)-(1-\tilde k')\mathrm{sn}(u\,|\,\tilde k^2)\mathrm{sd}(u\,|\,\tilde k^2),
\end{eqnarray}
where
\be\label{ktilde}
 k=\frac{1-\tilde k'}{1+\tilde k'}\;\Leftrightarrow\; \tilde k^2=\frac{4k}{(1+k)^2}
 \ee
 and the relations of the parameters with the new modulus \(\tilde k\) are
\begin{equation}
 \kappa=\frac{1-\tilde k'}{1+\tilde k'}\omega\;\Longrightarrow \tilde k'=\frac{\omega-\kappa}{\omega+\kappa},\qquad 
 \tilde k^2=\frac{4\kappa\omega}{(\omega+\kappa)^2}.
\end{equation}
Rescaling then the variable
\be
y=\frac{\omega\sigma}{1+\tilde k'}=\frac{\td \KK\,\s}{\pi}
\ee
 and exploiting the relations
 \be\label{usefulelliptic1}
 \KK(k^2)=\frac{1+\tilde k'}{2}\KK(\tilde k^2),\qquad
 \EE(k^2)=\frac{1}{1+\tilde k'}\left(\EE(\tilde k^2)+\tilde k'\KK(\tilde k^2)\right)
\ee
and
\begin{equation}\label{usefulelliptic2}
 \frac{\EE(k^2)}{\KK(k^2)}=\frac{2}{(1+\tilde k')^2}\Big[\frac{\EE(\tilde k^2)}{\KK(\tilde k^2)}+\tilde k'^2\Big]\ , 
\end{equation}
one obtains  two single-gap  Lam\'e operators 
 \ba\label{psi+bis}
\O_{\psi_+}&=& -\d^2_y+2\,\td {k}^2
  \sn^2(y+\frac{\widetilde{\KK}}{2}\,|\,\td {k}^2)-\td {k}^2+\frac{\pi^2\,\Om^2}{\widetilde{\KK}^2} \\
  \label{psi-bis}
 \O_{\psi_-}&=&-\d^2_y+2\,\td {k}^2
  \sn^2(y+\frac{3\,\widetilde{\KK}}{2})\,|\,\td {k}^2)-\td {k}^2+\frac{\pi^2\,\Om^2}{\widetilde{\KK}^2} 
\ea 
where the new elliptic parameter is  $\td{k}^2=\frac{4\,k}{(1+k)^2}$,  $\widetilde{\KK}=\KK(\td k^2)$, and the periodicity  $ \psi_\pm(y)=\psi_\pm(y+2\td \KK)$ is in the new variable $y=\frac{\td \KK}{\pi}\s$.

\end{description}


\appendix
\refstepcounter{section}
\def\theequation{B.\arabic{equation}}
\setcounter{equation}{0}
\section*{Appendix B: Determinant via Gel'fand-Yaglom method}

 For a periodic potential, we can compute the determinant via the discriminant as in (\ref{hill2}). In certain special cases, the discriminant can be found exactly because we know the explicit solutions $f_{1,2}$ in (\ref{norm}). This is the case, for example, for the constant potential $V(x)=m^2$, and also for the single-gap Lam\'e potential $V(x)=2k^2\,{\rm sn}^2(x | k^2)$. But in our comparison between the static gauge and conformal gauge we will need the determinant for {\it coupled} operators, and here we can use the Gel'fand-Yaglom theorem. To introduce this, we first state it for uncoupled operators, and then for coupled operators.
 
For the uncoupled equation (\ref{secondorder}) with periodic or antiperiodic boundary
 conditions, the Gel'fand Yaglom method \cite{dunne,kleinert} is simply an evaluation of the discriminant. That is, we numerically [or analytically, in special cases] evaluate the discriminant, using initial value boundary conditions (\ref{norm}) for the two independent functions $f_{1,2}(x; \Lambda)$. Note that here $\Lambda$ is just a parameter, so we are solving a homogeneous problem, with initial value conditions, which is numerically trivial. In fact, we can specify the initial conditions at any arbitrary point $\bar x$, and simply evolve through one period to evaluate the discriminant. That is, we can take initial conditions
\ba\label{GYbc}
f_1(\bar x; \Lambda)=1 \qquad &;& \qquad f^\prime_1(\bar x; \Lambda)=0\nonumber\\
f_2(\bar x; \Lambda)=0 \qquad &;& \qquad f^\prime_2(\bar x; \Lambda)=1
\ea
Then the Gelf'and-Yaglom theorem states that the determinant  with period $L$ is
\ba\label{detP}
\det_{P}^{\rm GY}(\Lambda) &=&\Delta(\Lambda)-2\nonumber\\
&=&f_1(\bar x+L; \Lambda)+f_2'(\bar x+L; \Lambda)-2
\ea
We can illustrate this method for the single-gap Lam\'e system, by taking two linear combinations of $f_+$ and $f_-$ in (\ref{solperiodic}) 
\be\label{f12}
f_1(x)=m_1\,f_+(x)+m_2\,f_-(x)\quad , \quad f_2(x)=n_1\,f_+(x)+n_2\,f_-(x)
\ee
such that the (\ref{GYbc}) are satisfied and $\a$ is given by  (\ref{alphaeq}).
One finds
\be\label{mn}
m_1=-\frac{f'_-(\bar x)}{D(\bar x)}\quad , \quad m_1=\frac{f'_+(\bar x)}{D(\bar x)}\quad ; \quad n_1=\frac{f_-(\bar x)}{D(\bar x)}\quad , \quad n_2=-\frac{f_+(\bar x)}{D(\bar x)}
\ee
where
\be\label{D}
D(\bar x)=f'_+(\bar x)\,f_-(\bar x)-f'_-(\bar x)\,f_+(\bar x)
\ee
Exploiting the monodromy
\be
f_{\pm}(x+2\KK)=-f_{\pm}(x)\,e^{\mp\,2\,\KK\, Z(\alpha | k^2)}\quad , \quad f'_{\pm}(x+2\KK)=-f'_{\pm}(x)\,e^{\mp\,2\,\KK\, Z(\alpha | k^2)},
\ee
it is then easy to check that the expression for the determinant (\ref{detP}) yields
\be\label{detP2}
{\rm Det}_{\rm P}=2\cosh[4\KK\,Z(\alpha | k^2)]-2 = 4\sinh^2\big[2\KK\,Z(\alpha | k^2)\big]
\ee
as before  (\ref{periodicdet}). Note the important observation that this result is \emph{independent of the initial point $\bar x$}. 
A specific example of the linear combinations (\ref{f12}) satisfying (\ref{GYbc}) at the initial point $\bar x=0$ are the solutions~\cite{BradenPeriodic}
\be\label{chi12}
f_1(x)= f_+(x)-f_-(x)\quad , \quad f_2(x)=\frac{\sn(\a\,|\,k^2)}{\cn(\a\,|\,k^2)\,\dn(\a\,|\,k^2)}\,\Big(f_+(x)+f_-(x)\Big).
\ee

For a system of coupled equations, the Gel'fand-Yaglom theorem generalizes in a straightforward manner. Consider (\ref{secondorder}) with $V(x)$ now an $n\times n$ matrix, and $f(x)$ an $n$-component column vector. Then we define $2n$ independent solutions $f_1^{(a)}(x; \Lambda)$ and  $f_2^{(a)}(x; \Lambda)$, for $a=1, 2, \dots n$, with the initial conditions expressed as a $2n\times 2n$ matrix:
\begin{eqnarray}
\begin{pmatrix}
f_1^{(1)}(0; \Lambda) &\dots & f_1^{(n)}(0; \Lambda) & f_1^{(1)\, \prime}(0; \Lambda) &\dots & f_1^{(n)\, \prime}(0; \Lambda)  \cr
f_2^{(1)}(0; \Lambda) &\dots & f_2^{(n)}(0; \Lambda) & f_2^{(1)\, \prime}(0; \Lambda) &\dots & f_2^{(n)\, \prime}(0; \Lambda)  
\end{pmatrix}={\mathbb I}_{2n\times 2n}
\end{eqnarray}
Then the Gel'fand-Yaglom theorem states that the infinite dimensional determinant can be expressed as a finite dimensional determinant:
\begin{eqnarray}
&&\det_P\left[-\del_x^2+V(x)-\Lambda\right]=\\\no
&&= -\det_{2n\times 2n}\left[{\mathbb I}-
\begin{pmatrix}
f_1^{(1)}(L; \Lambda) &\dots & f_1^{(n)}(L; \Lambda) & f_1^{(1)\, \prime}(L; \Lambda) &\dots & f_1^{(n)\, \prime}(L; \Lambda)  \cr
f_2^{(1)}(L; \Lambda) &\dots & f_2^{(n)}(L; \Lambda) & f_2^{(1)\, \prime}(L; \Lambda) &\dots & f_2^{(n)\, \prime}(L; \Lambda)  
\end{pmatrix}\right]
\end{eqnarray}
Again, this is completely straightforward to evaluate numerically. This was used to evaluate the determinant of the coupled operators in the conformal gauge example discussed in Section 5.


\appendix
\refstepcounter{section}
\def\theequation{C.\arabic{equation}}
\setcounter{equation}{0}

\def \te {\textstyle} 

\section*{Appendix C: Relevant expansions of elliptic and Jacobi functions}

\subsection*{Expansions of  ${\mathbb K}$ and ${\mathbb E}$ ($k\to1$)}

The expansion of the complete elliptic integrals in \(k'^2\) for \(k\to 1\) reads as follows  (with \(L=\ln4/k'\))
\begin{eqnarray}
 \!\!\!\!\!\! \KK(k^2)&=&\te{ L+\frac{1}{4}\left(L-1\right)k'^2+\frac{9}{64}\left(L-
 \frac{7}{6}\right)k'^4+\frac{25}{256}\left(L-\frac{37}{30}\right)k'^6+...}\\\no
\!\!\!\!\!\! \EE(k^2)&=& \te{1+\frac{1}{2}\left(L-\frac{1}{2}\right)k'^2+\frac{3}{16}\left(L
 -\frac{13}{12}\right)k'^4+\frac{15}{128}\left(L-\frac{6}{5}\right)k'^6} \\
 &&\ \ \ \ \ \ \ \ +\ \te{ \frac{175}{2048}\left(L
 -\frac{1051}{840}\right)k'^8+...}
\end{eqnarray}
Also
\ba\no
 \frac{\EE(k^2)}{\KK(k^2)}&=&\frac{1}{L}+\te{ \left(\frac{1}{2}-\frac{1}{2L}+\frac{1}{4L^2}\right)k'^2+
 \left(\frac{1}{16}-\frac{3}{32L}-\frac{3}{128L^2}+\frac{1}{16L^3}\right)k'^4}\\ &&+\te{ \left(\frac{1}{32}-\frac{3}{64L}-\frac{11}{768L^2}+\frac{5}{256L^3}+\frac{1}{64L^4}\right)k'^6+...}
\ea
This gives
\begin{equation}
 \kappa=\te{ \kappa_0-\frac{1}{4\pi}(\pi\kappa_0-2)\eta+\frac{9}{64\pi}\left(\pi\kappa_0-\frac{7}{3}\right)\eta^2
 -\frac{25}{256\pi}\left(\pi\kappa_0-\frac{37}{15}\right)\eta^3+...}
\end{equation}

\

\subsection*{Expansion of Jacobi Zeta function $Z$ ($k\to1$)}

Consider the integral representation of the Jacobi Zeta function (\ref{intzeta})
\begin{equation}\label{JacobiInt}
 f(y)=Z(\mathrm{sn}^{-1}(y|k^2)\,|\,k^2)
\end{equation}
We find for the asymptotics for \(k\to 1\) (setting \(L=\ln\frac{4}{k'}\))
\begin{eqnarray}
&&f(y)= y-\frac{1}{L}\int_0^y\frac{\mathrm{d}t}{1-t^2}-\frac{1}{2}k'^2y+
 \frac{1}{2L}k'^2\int_0^y\frac{\mathrm{d}t}{(1-t^2)^2}-\frac{1}{4L^2}k'^2\int_0^y\frac{\mathrm{d}t}{1-t^2}+ \\\no
&&~~+ k'^4\int_0^y\mathrm{d}t\left[-\frac{1}{16}\frac{1-5t^2+2t^4}{(1-t^2)^2}-\frac{1}{32L}\frac{-3+14t^2+t^4}{(1-t^2)^3}
 +\frac{1}{128L^2}\frac{3+13t^2}{(1-t^2)^2}-\frac{1}{16L^3}\frac{1}{1-t^2}\right]
\end{eqnarray}
which gives
\begin{eqnarray}
&&\!\!\!\!\!\!\!\!\!\!\!\!
f(y)= y-\frac{1}{L}\mathrm{artanh}(y)+\frac{1}{2}k'^2\left[-y+\frac{1}{2L}(\frac{y}{1-y^2}+\mathrm{artanh}(y))-
 \frac{1}{2L^2}\mathrm{artanh}(y)\right]\\\no &&\!\!\!\!\!\!\!\!\!\!\!\!+\frac{1}{16}k'^4\left[-\frac{y(1-2y^2)}{1-y^2}+\frac{1}{4L}\left(\frac{y(1-7y^2)}{(1-y^2)^2}+5\mathrm{artanh}(y)\right)+ \frac{1}{8L^2}\left(\frac{8y}{1-y^2}-5\mathrm{artanh}(y)\right)\right.\\\no
 &&\left.\!\!\!\!\!\!\!\!\!\!\!\! -\frac{1}{L^3}\mathrm{artanh}(y)\right]+\frac{1}{32}k'^6\left[-\frac{y(1-2y^2+2y^4)}{(1-y^2)^2}+\frac{1}{8L}\left(\frac{y(3-20y^2+57y^4)}{3(1-y^2)^3}+
 11\mathrm{artanh}(y)\right)\right.\nonumber\\\no
 & &\!\!\!\!\!\!\!\!\!\!\!\! \left.+\frac{1}{48L^2}\left(\frac{9y(5-9y^2)}{(1-y^2)^2}-23\mathrm{artanh}(y)\right)-\frac{1}{8L^3} \left(\frac{4y}{1-y^2}-9\mathrm{artanh}(y)\right)-\frac{1}{2L^4}\mathrm{artanh}(y)\right]+...
\end{eqnarray}
Using the results above, one can read off the expansions of the relevant quantities appearing in the static gauge fluctuation determinants 
\ba\no
&&\!\!\!\!\!\!\!\!\! 2\KK(k^2)\,Z(\a_\b\,|\,k^2)\sim \k_0 \pi  x-2 \tanh ^{-1}x +\eta\,\frac{\left(\k_0 \pi  x^2-x^2-\k_0 \pi +2\right) }{\k_0 \pi  x \left(x^2-1\right)}+\\\no
&&+\,\frac{\eta ^2}{32 \k_0^2 \pi
   ^2 x^3 \left(x^2-1\right)^2}\,\Big[\k_0^3 \pi ^3 x^6-13 \k_0^2 \pi ^2 x^6+21 \k_0
   \pi  x^6-4 \k_0^3 \pi ^3 x^4+30 \k_0^2 \pi ^2 x^4+\\\label{2KZbeta}
   &&-71 \k_0 \pi  x^4+32 x^4+3 \k_0^3 \pi ^3
   x^2-19 \k_0^2 \pi ^2 x^2+66 \k_0 \pi  x^2-80 x^2-16 \k_0 \pi +32 \Big]+...\\\no
   &&\\\no
&&\!\!\!\!\!\!\!\!\! 2\KK(\td{k}^2)\,Z(\a_\p\,|\,\td{k^2})\sim\k_0 \pi  y-2 \tanh ^{-1} \frac{y}{2}+\eta\,\frac{2 (\k_0 \pi -1)}{\k_0 \pi 
   y}+\,\frac{\eta^2}{16 \pi ^2 y^3 \omega
   ^2}\, \big(\k_0^3 \pi ^3 y^4-13 \k_0^2 \pi ^2 y^4+\\\label{2KZphi}
   &&
   +21 \k_0 \pi  y^4-4 \k_0^3 \pi ^3 y^2+56
   \k_0^2 \pi ^2 y^2-116 \k_0 \pi  y^2+32 y^2+128 \k_0 \pi -128\big)+...\\\no
   &&\\ \no
&&\!\!\!\!\!\!\!\!\!\KK(\bar\nu)\,Z(\td\a)\sim \k_0 \pi  z-\tanh ^{-1}z +\eta\,\frac{(\k_0 \pi -1) }{2 \k_0 \pi 
   z}+\,\frac{\eta ^2}{64 \k_0^2 \pi ^2 z^3
   \left(z^2-1\right)}\,\big(\k_0^3 \pi ^3 z^4-13 \k_0^2 \pi ^2 z^4+\\\label{2KZpsi}
   && +21 \k_0 \pi  z^4-\k_0^3 \pi ^3 z^2+14
   \k_0^2 \pi ^2 z^2-29 \k_0 \pi  z^2+8 z^2+8 \k_0 \pi -8\big) +...
\ea
where $x,y,z$ are defined in (\ref{x})-(\ref{y})-(\ref{z}).

\subsection*{Expansions of  ${\mathbb K}$ and ${\mathbb E}$ ($k\to0$)}

The first few orders of the $\k\to0$ expansions for the elliptic integrals  $\KK$ and $\E$ read
\ba
\KK(k^2)&=&\frac{\pi }{2}+\frac{\pi  }{8}k^2+\frac{9 \pi }{128} k^4+\frac{25 \pi 
   }{512}k^6+\frac{1225 \pi }{32768} k^8+O\left(k^9\right)\\
   \EE(k^2)&=&\frac{\pi }{2}-\frac{\pi  }{8}k^2-\frac{3 \pi  }{128}k^4-\frac{5 \pi 
  }{512} k^6-\frac{175 \pi  }{32768}k^8+O\left(k^9\right)
\ea

\subsection*{Expansion of Jacobi Zeta function $Z$ ($k\to0$)}

As  efficient way to evaluate the small  spin expansion ($k\to0,~\eta\to\infty$) presented in  Section 7 is to  
first compute the expansion of $\d Z(\alpha\,|\,k^2)/\d\Omega$, where the dependence of $Z$ on $\Omega$ is via $\alpha$ as 
solution of the (\ref{detbeta2})-(\ref{detpsi2}), and then perform an indefinite integration over $\Omega$.

Using (\ref{zed}) valid for $0<\alpha<\KK$, after some straightforward manipulations one can write
\be\label{derZ}
\frac{\d\,{Z}(\a\,|\,k^2)}{\d\Om}=\frac{1}{\sqrt{1-\sn^2(\a\,|\,k^2)}\,\sqrt{1-k^2\,\sn^2(\a\,|\,k^2)}}\,
\frac{\d\,\sn(\a\,|\,k^2)}{\d\Om}\,\Big[1-k^2\,\sn(\a\,|\,k^2)-\frac{\EE}{\KK}\Big]
\ee

Considering the determinant for beta modes
\be
\det\,\O_\beta=4\sinh^2\left[2\KK Z(\a|k^2)\right],\qquad \sn(\a|k^2)=\frac{1}{k}\sqrt{1+k^2+\left(
\frac{\pi\Om}{2\KK}\right)^2}
\ee
one can see, that \(\sn(\a\,|\,k^2)>1\). So 
before proceeding with the short string
expansion, one needs the following transformation:
\be
 \alpha=\beta+\KK+i\KK'
\ee
which gives
\be\label{eq:newbeta}
 \sn(\b|k^2)=\sqrt{\frac{k^2+\left(\frac{\pi\Omega}{2\KK}\right)^2}{1+\left(\frac{\pi\Omega}{2\KK}\right)^2}}<1
\ee
This affects the determinant in the following way:
\be
\det\,\O_\beta=4\sinh^2\left[2\KK Z(\beta|k^2)-2\KK
\frac{\sn(\beta|k^2)\dn(\beta|k^2)}{\cn(\beta|k^2)}-i\pi\right]
\ee
Substituting the explicit expression for $\sn(\beta\,|\,k^2)$ given by (\ref{eq:newbeta}) into (\ref{derZ}), expanding then for large $\eta$ and integrating 
back in $\Omega$, 
one can easily evaluate the expansions for the relevant determinants.
In the case of the fluctuation $\beta$  it is finally
\ba\no
\det\,\O_\beta\,&\equiv&\,4\sinh^2[2\,\KK Z(\alpha_\beta\,|\,k^2)]\,=\,
4 \sinh ^2(\pi  \Om )+\frac{1}{\eta}\,\frac{2 \pi  \sinh (2 \pi  \Om )}{  \Om }+\\\label{detbetashort}
&+&\frac{1}{\eta^2}\,\Big[\frac{\pi ^2 \cosh (2 \pi \Om )}{\Om ^2}-\frac{\pi  \left(3 \Om ^4
+6 \Om ^2+2\right) \sinh (2 \pi  \Om )}{4 \left(\Om ^5+\Om ^3\right)}\Big]+O\left(\frac{1}{\eta^3 }\right)~.
\ea
Applying this approach to the fermion determinant
\be
 \det\,\O_\psi=-4\cosh^2\left[\tilde\KK Z(\alpha|\tilde k^2)\right],\qquad \sn(\alpha|\tilde k^2)=\frac{1}{\tilde k}
 \sqrt{1+\left(\frac{\pi\Om}{\tilde\KK}\right)^2}
\ee
by using \(\alpha=\beta+\tilde\KK+i\tilde\KK'\) with
\be
 \sn(\beta|\tilde k^2)=\sqrt{\frac{\left(\frac{\pi \Om}{\tilde\KK}\right)^2}{1-\tilde k^2+\left(\frac{\pi\Om}{\tilde\KK}
 \right)^2}}
\ee
gives
\ba
 \det\,\O_{\psi}&=&-4\cosh^2\left[\tilde\KK Z(\beta|\tilde k^2)-\tilde\KK\,
\frac{\sn(\beta|\tilde k^2)\dn(\beta|\tilde k^2)}{\cn(\beta|\tilde k^2)}-\frac{i\pi}{2}\right]
\nonumber\\
&=&4\sinh^2\left[\mathbb{\tilde K}Z(\beta|\tilde k^2)-\mathbb{\tilde K}
\frac{\sn(\beta|\tilde k^2)\dn(\beta|\tilde k^2)}{\cn(\beta|\tilde k^2)}\right],
\ea
where the term \(i\pi/2\) flips the \(\cosh\) to \(\sinh\).The expansion of this expression in \(1/\eta\) gives
\be
 \det\,\O_{\psi}(\Om)=D_{\psi}^{(0)}(\Om)+\frac{1}{\eta}D_{\psi}^{(1)}(\Om)+\frac{1}{\eta^2}D_{\psi}^{(2)}(\Om)+...
\ee
with
\ba
 D_{\psi}^{(0)}(\Om)&=&4 \sinh(\pi \Om),\\
 D_{\psi}^{(1)}(\Om)&=&\frac{4 \pi \Om \sinh(2\pi \Om)}{1+4 \Om^2},\\
 D_{\psi}^{(2)}(\Om)&=&\frac{4 \pi ^2 \Om ^2 \cosh(2 \pi \Om)}{(1+4 \Om ^2)^2}-\frac{\pi \Om(48 \Om ^2(1+\Om ^2)+1)\sinh(2 \pi \Om)}{2
 (1+4 \Omega ^2)^3}
\ea


\appendix
\refstepcounter{section}
\def\theequation{D.\arabic{equation}}
\setcounter{equation}{0}

\section*{Appendix D: Exponentially suppressed contributions}

As explained below (\ref{ratiocoth3}) and around (\ref{expcorr}), in performing the large spin expansion on the exact determinants we systematically adopted an approximation based on the replacement $\tanh(\cdots)\to 1$. The neglected terms 
are exponential in the large quantity $\kappa_0$ and give back powers of $\eta$, see (\ref{expcorr}). Lacking a better complete control of this approximation, 
we present in this Appendix the evaluation of the leading large $\kappa_0$ correction to the one-loop energy due to 
the above replacement.
 It is clear that such leading correction come indeed from the fermion determinant, that  has the following leading order in the formal small $\eta$, {\em i.e.} large spin, expansion
\be
{\cal D}_{\psi, \rm LO} = \frac{1}{1-z^2}\,\left[1-z\,\tanh(\pi\,\kappa_0\,z)\right]^2,\qquad z = \sqrt{1+\frac{\Omega^2}{\kappa_0^2}}.
\ee
and which we treated in the following approximated way
\be
{\cal D}_{\psi, \rm LO}^{\rm approx} = \frac{1}{1-z^2}\,(1-z)^2 = \frac{1-z}{1+z}.
\ee
The effect of this approximation in the one-loop energy is 
\be
\Delta E_1 = -\frac{1}{4\,\pi\,\kappa}\,\cdot 8\cdot \,\int_{-\infty}^\infty d\Omega\, (\ln{\cal D}_{\psi, \rm LO}-\ln{\cal D}_{\psi, \rm LO}^{\rm approx}) = 
-\frac{4\,\kappa_0}{\pi\,\kappa}\int_0^\infty d\overline\Omega\, F(\overline\Omega; \kappa_0),
\ee
where
\be
F(\overline\Omega; \kappa_0) = \ln\left(
\frac{1-\sqrt{1+\overline\Omega^2}\,\tanh(\pi\,\kappa_0\,\sqrt{1+\overline\Omega^2})}{1-\sqrt{1+\overline\Omega^2}}\right)^2.
\ee
The function $F(\overline\Omega; \kappa_0)$ has the generic shape shown in Fig.~(\ref{fig:F}).
The point where it goes to $-\infty$ is where the numerator inside the logarithm vanishes. This happens at approximately
\be
\overline\Omega^* \simeq 2\, e^{-\pi\,\kappa_0}.
\ee
The large $\kappa_0$ analysis must be done carefully since the two regions $\overline\Omega < \overline \Omega^*$ and 
$\overline\Omega > \overline \Omega^*$ contribute the above integrals with opposite signs and large cancellations.

\medskip

\begin{figure}[htb]
\begin{center}
\includegraphics[scale=0.8]{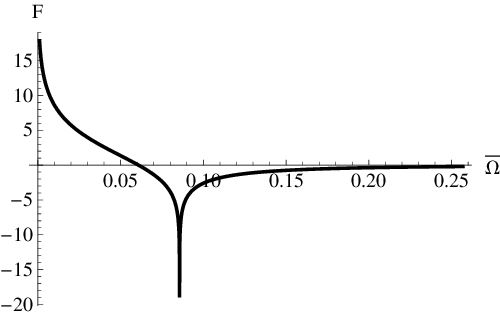}
\caption{Shape of the function $F(\overline\Omega; \kappa_0)$ at $\kappa_0 = 1$.}
\label{fig:F}
\end{center}
\end{figure}

\begin{figure}[htb]
\begin{center}
\includegraphics[scale=0.8]{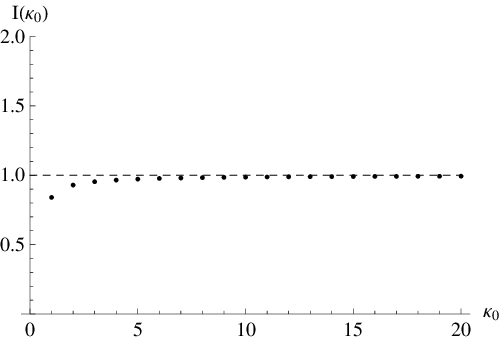}
\caption{Numerical evaluation of the 
integral $I(\kappa_0) = \frac{1}{8\,\pi\,\sqrt{\kappa_0}}\,e^{2\pi\kappa_0}\int_0^\infty d\overline\Omega\, F(\overline\Omega; \kappa_0)$ and comparison
with the leading analytic prediction 1 (dashed line).}
\label{fig:Fint}
\end{center}
\end{figure}

\begin{figure}[htb]
\begin{center}
\includegraphics[scale=0.8]{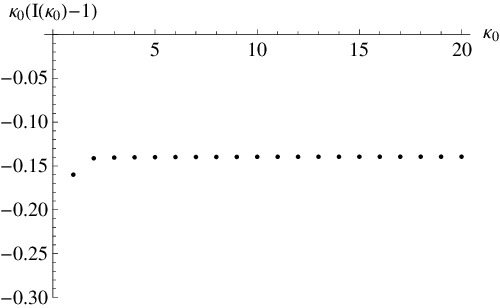}
\caption{Numerical evaluation of the quantity $\kappa_0(I(\kappa_0)-1)$.}
\label{fig:Fintbis}
\end{center}
\end{figure}

The final result is the estimate
\be
\int_0^\infty d\overline\Omega\, F(\overline\Omega; \kappa_0) = 8\,\pi\,\sqrt{\kappa_0}\,e^{-2\,\pi\,\kappa_0} + \cdots~,
\ee
whose accuracy is shown in Fig.~(\ref{fig:Fint}). In terms of the one-loop energy, it is 
\be
\Delta E_1 = -32\,\frac{\kappa_0^{3/2}}{\kappa}\,e^{-2\,\pi\,\kappa_0}  + \cdots~  = {\cal O}(\eta^2\, \ln^{1/2}\eta).
\ee
The peculiar half-integer exponent of $\ln\eta$ suggest that a systematic resummation of these corrections is needed possibly taking into account mixing with similar terms coming from the expansion of Bessel functions $K_1$ in Eqs.~(\ref{e0bessel})-(\ref{e0bessel1}). We did not evaluate the next correction, but 
Fig.~(\ref{fig:Fintbis}) suggest that 
\be
\int_0^\infty d\overline\Omega\, F(\overline\Omega; \kappa_0) = \left(8\,\pi\,\sqrt{\kappa_0} + \frac{a}{\sqrt\kappa_0} + \cdots\right)\, e^{-2\,\pi\,\kappa_0} + \cdots~,
\ee
for some constant $a$. This term has the same form as the Bessel $K_1$ corrections.


\appendix
\refstepcounter{section}
\def\theequation{E.\arabic{equation}}
\setcounter{equation}{0}
\section*{Appendix E: Large spin limit with  a different approximation}

The analysis of the large spin limit in the Section \ref{sec:large}, leading to the large spin limit expansion (\ref{ordereta3}), agrees extremely well with the exact numerical dependence of the one-loop energy on the elliptic parameter $k$, as can be seen from Figure \ref{exactVSnumerical}. However, this expansion neglected exponential terms in the $k\to 1$ limit, as explained in Section \ref{sec:leading}, and the significance of these neglected terms at smaller values of $k$ is not clear. In this section we consider another type of approximation, that leads to a different type of large spin expansion of the one-loop energy.

Start from the effective action 
\be
\Gamma = -\frac{\cal T}{4\,\pi}\,\int_{\mathbb{R}} d\Omega\, \log\frac{\det^8{\cal O}_\psi}{\det {\cal O}_\phi\,\det^2{\cal O}_\beta\,\det^5{\cal O}_0},
\ee
and integrate by parts (we set also ${\cal T}=1$)
\ba
\Gamma &=& \frac{1}{4\,\pi}\,\int_{\mathbb{R}} d\Omega\,\Omega\,\partial_\Omega\left[8\,\log\det{\cal O}_\psi-\log\det{\cal O}_\phi-2\,\log\det{\cal O}_\beta-5\,\log\det{\cal O}_0\right]  \nonumber\\
&\equiv& 8\,\Gamma_\psi-\Gamma_\phi-2\,\Gamma_\beta-5\,\Gamma_0.
\ea
We now systematically apply the approximate substitutions
\be
\log(4\,\sinh^2 x) \to 2 |x|\quad ,\qquad\log(4\,\cosh^2 x) \to 2 x \quad,
\ee
which also correspond to neglecting exponential terms, but now the approximation is made {\bf before} integrating over $\Omega$.
Then, the effective action can be computed exactly, {\em i.e.} without any leftover integral. As an example,
let us consider the contribution from the $\phi$ mode. We convert the integral over $\Omega$ into an integral over the spectral parameter $\alpha_\phi$ as follows:
\ba
\Gamma_\phi &=& \frac{1}{4\,\pi}\int_{-\infty}^\infty d\Omega\,\Omega\,\partial_\Omega\,\log(4\,\sinh^2(2\,\widetilde{\mathbb{K}}\,Z(\alpha_\phi\,|\,\widetilde k^2))\nonumber\\
& \simeq& \frac{2\,\widetilde{\mathbb{K}}}{\pi} \int_0^\infty d\Omega\,\Omega\,\partial_\Omega Z(\alpha_\phi\,|\,\widetilde k^2)) \nonumber\\
&=&
 \frac{2\,\widetilde{\mathbb{K}}}{\pi} \int_{\alpha_0}^{i\,\widetilde{\mathbb{K}}'} d\alpha\,\Omega\,\partial_\alpha Z(\alpha_\phi\,|\,\widetilde k^2) \nonumber \\
&=& \left(\frac{2\,\widetilde{\mathbb{K}}}{\pi}\right)^2 \int_{\alpha_0}^{i\,\widetilde{\mathbb{K}}'} d\alpha\,\widetilde\Omega\,\left[{\rm dn}^2(\alpha\,|\,\widetilde k^2)-\frac{\widetilde{\mathbb{E}}}{\widetilde{\mathbb{K}}}\right],
\ea
where 
\be
\widetilde\Omega = \sqrt{-{\rm dn}^2(\alpha\,|\,\widetilde k^2)},
\ee
and $\alpha_{0}$ is the value associated with $\Omega=0$. The spectral parameter $\alpha_\phi$ takes values along the straight line joining $\widetilde{\mathbb K}+i\widetilde {\mathbb K}^\prime$ to $i\widetilde {\mathbb K}^\prime$, so it is convenient to write
\be
\alpha = \widetilde{\mathbb{K}}+i\,\widetilde{\mathbb{K}}'-\beta,
\ee
and use the identity
\be
{\rm dn}^2( \widetilde{\mathbb{K}}+i\,\widetilde{\mathbb{K}}'-\beta\,|\,\widetilde k^2) = (\widetilde k^2-1)\,{\rm sc}^2(\beta\,|\,\widetilde k^2).
\ee
We find 
\ba
\Gamma_\phi 
&\simeq& \widetilde k'\,\left(\frac{2\,\widetilde{\mathbb{K}}}{\pi}\right)^2 \int_0^{\widetilde{\mathbb{K}}} d\beta\,{\rm sc}(\beta\,|\,\widetilde k^2)\,
\left[\frac{\widetilde{\mathbb{E}}}{\widetilde{\mathbb{K}}}+\widetilde k^2{}'\,{\rm sc}^2(\beta\,|\,\widetilde k^2)\right].
\ea
A similar treatment leads to
\be
\Gamma_\psi \simeq \frac{1}{4}\,\Gamma_\phi \quad ,
\ee
within this approximation,  and the following contribution from the $\beta$ modes
\ba
\Gamma_\beta &\simeq& \left(\frac{2\,{\mathbb{K}}}{\pi}\right)^2 \int_{{\rm sn}^{-1}(k\,|\,k^2)}^{{\mathbb{K}}} d\beta\,\sqrt{(1-k^2)\,{\rm nc}^2(\beta\,|\,k^2)-1}\,
\left[\frac{\mathbb{E}}{\mathbb{K}}+(1-k^2)\,{\rm sc}^2(\beta\,|\, k^2)\right].
\ea
It is convenient to introduce the variables
\be
s_{\phi} = {\rm sn}(\beta\,|\,\widetilde k^{2}),\qquad 
s_{\beta} = {\rm sn}(\beta\,|\, k^{2}).
\ee
The integrals can be computed in closed-form, and after a long calculation, one finds 
\ba
\Gamma_\phi  &=& -\frac{(k-1)^2}{\pi^2}\,\mathbb{K}^2 \,\frac{1}{s_{\phi}-1}+\frac{4(\mathbb{K}-\mathbb{E})\mathbb{K}}{\pi^2}\,\log(1-s_{\phi}) + \Gamma_\phi^{\rm finite}, \\
\Gamma_\beta &=& \frac{k^2-1}{\pi^2}\,\mathbb{K}^2 \,\frac{1}{s_{\beta}-1}+\frac{2(\mathbb{K}-\mathbb{E})\mathbb{K}}{\pi^2}\,\log(1-s_{\beta}) + \Gamma_\beta^{\rm finite}.
\ea
The integrals are divergent at $s_{\phi}, s_{\beta}\to 1$, which is simply the individual UV divergence. Introducing
a cut-off $\Omega_{\rm max}$ with 
\ba
s_{\beta, \rm max} &=& 1-\frac{2\,\mathbb{K}^2(k^2)}{\pi^2\,\Omega_{\rm max}^2}\,(1-k^2) + \cdots, \\
s_{\phi, \rm max}  &=& 1-\frac{2\,\widetilde{\mathbb{K}}^2(\widetilde k^2)}{\pi^2\,\Omega_{\rm max}^2}\,(1-\widetilde k^2) + \cdots, \\
s_{\psi, \rm max} &=& 1-\frac{\widetilde{\mathbb{K}}^2(k^2)}{2\,\pi^2\,\Omega_{\rm max}^2}\,(1-\widetilde k^2) + \cdots.
\ea
one checks that the pole cancels
against the free field contribution which is 
\ba
\Gamma_{0} &=& \frac{1}{4\,\pi}\int_{-\Omega_{\rm max}}^{\Omega_{\rm max}} d\Omega\,\Omega\,\partial_\Omega\,\log(4\,\sinh^2(\pi\,\Omega)) \nonumber\\
&\simeq & \int_0^{\Omega_{\rm max}} d\Omega\,\Omega = \frac{1}{2}\,\Omega^{2}_{\rm max.}
\ea
\begin{figure}[hbtp]
\begin{center}
\includegraphics[scale=0.6]{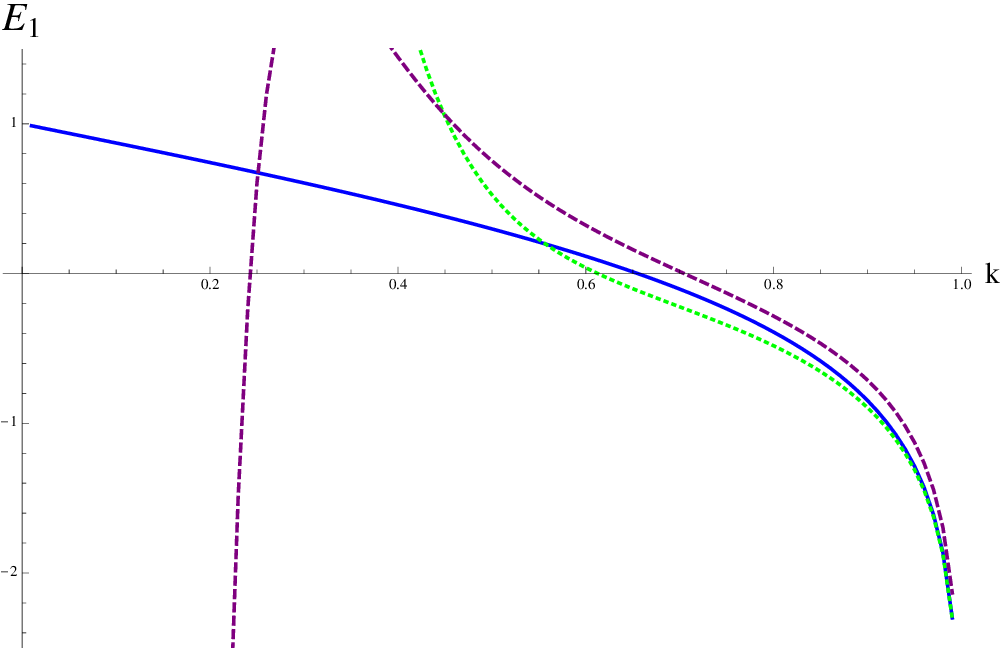}
\end{center}
\small
    \caption{Plots  of $E_1$ as a function of $k$: the blue (solid) curve is found numerically from the exact 
    expression \rf{final2} for generic  values of $k$;
 the green (dotted) curve is found from an analytic expansion in the 
  $k\to 1 $ or large  spin limit, using  the first two terms in (\ref{ordereta3}); the purple (dashed) curve is the first two terms in the alternative $k\to 1$ expansion (\ref{different}).
    }
\label{comparison-plot}
\end{figure}
The logarithmic divergence cancels in the sum of the various mode contributions.
The full result is quite compact and reads
\ba\no
\Gamma&\simeq &\frac{\KK(k^2)}{\pi ^2 \KK(k^4)} \Big[4\KK(k^2) \EE(k^4)\,{\rm Im}\, \mathbb{F}\Big(\arcsin\frac{1}{k}\,|\,k^4\Big) +\KK(k^2) \Big(\pi+2 \KK(k^4) \,(k^2-2 \ln k-1-6 \ln2)
   \Big)+\\
   &&   +4 \KK(k^4) \EE(k^2) \left(\ln(8 k)-(1+  k^2)\,{\rm Im}\, \mathbb{F}\Big(\arcsin\frac{1}{k}\,|\,k^4\Big)  \right)\Big].
\ea
Since this compact expression is written in closed-form in terms of elliptic functions, its $k\to 1$ expansion can be computed and the result to order ${\cal O}(\eta^{4})$ is
\ba
\Gamma &\simeq&
\kappa _0 \left(1+\frac{6 \log {2}}{\pi }\right)-3 \kappa _0^2 \log {2} + \nonumber\\
&& + \left[
\kappa _0 \left(-\frac{1}{\pi }-\frac{3 \log 2}{\pi }\right)+\frac{1}{2 \pi }+\frac{3 \log 2}{\pi ^2}
\right]\,\eta + \nonumber \\
&& + \left[
\kappa _0^2 \left(-\frac{\pi }{32}-\frac{3 \log 2}{32}\right)+\kappa _0 \left(\frac{1}{16}+\frac{1}{2 \pi }+\frac{39 \log 2}{32 \pi
   }\right)-\frac{13}{64 \pi }-\frac{1}{2 \pi ^2}-\frac{63 \log 2}{32 \pi ^2}
\right]\,\eta^2 + \nonumber\\
&& + \left[
\kappa _0^2 \left(\frac{1}{96}+\frac{\pi }{32}+\frac{3 \log 2}{32}\right)+\kappa _0 \left(-\frac{3}{32}-\frac{1}{3 \pi }-\frac{13
   \log 2}{16 \pi }\right)+\frac{29}{192 \pi }+\frac{29}{64 \pi ^2}+\frac{85 \log 2}{64 \pi ^2}
\right]\,\eta^3 + \nonumber\\
&& + \left[
\kappa _0^2 \left(-\frac{1}{64}-\frac{115 \pi }{4096}-\frac{693 \log 2}{8192}\right)+\kappa _0 \left(\frac{51}{512}+\frac{25}{96 \pi
   }+\frac{10263 \log 2}{16384 \pi }\right)+\right.\nonumber\\
   &&\left. -\frac{4397}{32768 \pi }-\frac{149}{384 \pi ^2}-\frac{16403 \log 2}{16384 \pi ^2}
   \right]\,\eta^4 + \cdots
   \label{different}
\ea
Recalling that $E_1=\Gamma/\kappa$, we see that this expansion is very similar to, but not precisely the same as, the $k\to 1$ expansion found  in (\ref{ordereta3}). A comparison of these two different approximations is presented in Figure \ref{comparison-plot}. Both agree with the exact result at large spin, but we see that the expansion in (\ref{ordereta3}) provides a better approximation as $k\to 1$. Nevertheless, the approximation considered in this appendix may be of interest as it provides a closed-form expression.


\appendix
\refstepcounter{section}
\def\theequation{F.\arabic{equation}}
\setcounter{equation}{0}

\section*{Appendix F: Static gauge determinants in perturbation theory}

We repeat here an evaluation of the determinants in the short string limit adopting the standard  perturbation theory method of~\cite{tt}.

The basic idea is to compute
\ba
\ln\det(-\partial^2+\Om^2+V) &=& \ln\det(-\partial^2+\Om^2) + 
{\rm Tr}\Big(\frac{1}{-\partial^2+\Om^2}\,V\Big) \\
&& -\frac{1}{2}
{\rm Tr}\Big(\frac{1}{-\partial^2+\Om^2}\,V\,\frac{1}{-\partial^2+\Om^2}\,V\Big)+\cdots~,\nonumber
\ea
evaluating the traces on the following basis $|n\rangle = \frac{1}{\sqrt{2\pi}}\,e^{i\,n\,\sigma}$, $n\in\mathbb{Z}$, $\sigma\in[0,2\pi]$.
In all cases, we define
\be
V_{n,m} = \frac{1}{2\,\pi}\int_0^{2\pi} d\sigma e^{i\,(m-n)\,\sigma}\,V(\sigma)
\ee
and also 
\be
\ln\det{\cal O}_f = \ln\det(-\partial^2+\Om^2) + X_f,\ \ \ \ 
\qquad f = (\beta, \phi, \psi).
\ee

\bigskip
\noindent
For the \emph{$\beta$ mode}, we have (see (\ref{lamebeta}) and the rescaling (\ref{betarescaled}))
\be
V_\beta = \Big(\frac{2\,\mathbb{K}(k^2)}{\pi}\Big)^2\,2\,k^2\,{\rm sn}^2(\frac{2\,\mathbb{K}(k^2)}{\pi}\,\sigma\,|\,k^2)
 = \frac{2}{\eta}\sin^2\sigma + \cdots,
\ee
and the non zero matrix elements
\be
V_{n,n} = \frac{1}{\eta}, \qquad V_{n, n\pm 2} = -\frac{1}{2\eta}.
\ee
Thus
\be
X_\beta = \frac{1}{\eta}\sum_{n\in\mathbb{Z}}\frac{1}{n^2+\Om^2}.
\ee

\bigskip
\noindent
For the \emph{$\phi$ mode}, we have (see (\ref{phibis}) and the rescaling (\ref{phirescaled})~\footnote{The imaginary shift   in the argument of the elliptic sine in (\ref{phibis}) is  irrelevant for the determinant calculation since it does not change the monodromy. For the check here proposed it is useful to consider the analytically continued potential (\ref{Vcontinued}).  })
\be\label{Vcontinued}
V_\phi =\Big(\frac{2\,\mathbb{K}(\td{k}^2)}{\pi}\Big)^2\,\Big[2\,k^2\,{\rm sn}^2(\frac{2\,\mathbb{K}(\td {k^2})}{\pi}\,\sigma\,|\,\td{k}^2)-\td{k}^2\Big]= -\frac{4}{\sqrt\eta}\cos(2\sigma)+\frac{4}{\eta}\sin^2(2\sigma)+\cdots~,
\ee
and the non zero matrix elements
\be
V_{n,n} = \frac{2}{\eta}, \qquad V_{n, n\pm 2} = \frac{2}{\sqrt\eta},
\qquad V_{n, n\pm 4} = -\frac{1}{\eta}.
\ee
Thus
\be
X_\phi = \frac{2}{\eta}\sum_{n\in\mathbb{Z}}\frac{1}{n^2+\Om^2}\left(1-\frac{1}{(n+2)^2+\Om^2}-\frac{1}{(n-2)^2+\Om^2}\right).
\ee

\bigskip
\noindent
For the \emph{$\psi$ mode} (see (\ref{psi+bis})-(\ref{psi-bis}) and (\ref{psirescaled})), we have 
\be
V_\psi = -\frac{1}{\sqrt\eta}\cos \sigma+\frac{1}{\eta}\sin^2\sigma+\cdots~,
\ee
and the non zero matrix elements
\be
V_{n,n} = \frac{1}{2\eta}, \qquad V_{n, n\pm 1} = -\frac{1}{2\sqrt\eta},
\qquad V_{n, n\pm 2} = -\frac{1}{4\eta}.
\ee
Thus
\be
X_\psi = \frac{1}{2\eta}\sum_{n\in\mathbb{Z}}\frac{1}{n^2+\Om^2}\left(1-\frac{1}{4}\frac{1}{(n+1)^2+\Om^2}-\frac{1}{4}\frac{1}{(n-1)^2+\Om^2}\right).
\ee

\bigskip
\noindent
For the combination, we have
\ba
8\,X_\psi-2\,X_\beta-X_\phi &=& 
-\frac{1}{\eta}\sum_{n\in\mathbb{Z}}\frac{1}{n^2+\Om^2}\left(\frac{1}{(n+1)^2+\Om^2}+\frac{1}{(n-1)^2+\Om^2}\right) + \nonumber \\
&& +\frac{2}{\eta}\sum_{n\in\mathbb{Z}}\frac{1}{n^2+\Om^2}\left(\frac{1}{(n+2)^2+\Om^2}+\frac{1}{(n-2)^2+\Om^2}\right).\nonumber
\ea
Evaluating the infinite sums over $n$, we find 
\be
8\,X_\psi-2\,X_\beta-X_\phi  = \frac{1}{\eta}\,\frac{2 \pi  \left(2 \Om ^2-1\right) \coth (\pi  \Om )}
{\Om  \left(\Om ^2+1\right) \left(4 \Om^2+1\right)}.
\ee
This is the same as \refeq{IR} showing that the old-fashioned way of calculation is in perfect agreement with the the procedure adopted in this paper. 
 
 \bigskip 
 
Comparing now this with the calculation in conformal gauge of~\cite{tt}, one can see that the difference is due to the second order 
contribution of the $1/\sqrt\eta$ term from the $\phi$ field. It is 
\be
\frac{2}{\eta}\sum_{n\in\mathbb{Z}}\frac{1}{n^2+\Om^2}\left(\frac{1}{(n+2)^2+\Om^2}+\frac{1}{(n-2)^2+\Om^2}\right) = \frac{2 \pi  \coth (\pi  \Om )}{\Om  \left(\Om ^2+1\right)}.
\ee
Now, reconsider Eq.~(3.32) of \cite{tt} which reads
\be
\int_\mathbb{R} d\Om\sum_{n\in\mathbb{Z}}\left[\frac{2}{n^2+\Om^2}-\frac{i\,\Om}{n^2+(\Om+i)^2}+\frac{i\,\Om}{n^2+(\Om-i)^2}\right] = 0,
\ee
upon doing a shift in $\Om$ in the last two terms. Actually, one could perform first the sum over modes of the above integrand thus getting
\be
\sum_{n\in\mathbb{Z}}\left[\frac{2}{n^2+\Om^2}-\frac{i\,\Om}{n^2+(\Om+i)^2}+\frac{i\,\Om}{n^2+(\Om-i)^2}\right] = \frac{2 \pi  \coth (\pi  \Om )}{\Om  \left(\Om ^2+1\right)} \,  
\ee
This means that avoiding the shifts in Eq.~(3.32) of \cite{tt} one recovers full equality with the static gauge.


 
\appendix
 \refstepcounter{section}
\def\theequation{F.\arabic{equation}}
\setcounter{equation}{0}

\subsection*{ }

\end{document}